\documentclass[reqno,a4paper]{amsart}  
\usepackage{amsmath, latexsym, amsthm, amsfonts,bm,amssymb,mathscinet} %
\usepackage{mathtools}
\usepackage{xcolor}
\usepackage[plainpages=false,pdfpagelabels]{hyperref}
 
\usepackage{url} %
\usepackage{enumitem}
\usepackage{mathtools} %
\usepackage{booktabs} %
\usepackage{caption} %
\usepackage{verbatim}

\usepackage{natbib} %
\bibpunct{(}{)}{;}{a}{}{,} %

\theoremstyle{plain}
\newtheorem{thm}{Theorem}[section]
\newtheorem{lemma}[thm]{Lemma}

\theoremstyle{remark}
\newtheorem{remark}[thm]{Remark}

\newcommand{\ig}{\operatorname{IG}}
\newcommand{\gam}{\operatorname{Gamma}}
\newcommand{\Exp}{\operatorname{Exp}}
\def\iid{\stackrel{\textrm{i.i.d.}}{\sim}}

\allowdisplaybreaks

\numberwithin{equation}{section}
\numberwithin{figure}{section} 
\numberwithin{table}{section}

\usepackage[foot]{amsaddr}

\usepackage{tikz,xcolor,hyperref}

\definecolor{lime}{HTML}{A6CE39}
\DeclareRobustCommand{\orcidicon}{%
	\begin{tikzpicture}
	\draw[lime, fill=lime] (0,0) 
	circle [radius=0.16] 
	node[white] {{\fontfamily{qag}\selectfont \tiny ID}};
	\draw[white, fill=white] (-0.0625,0.095) 
	circle [radius=0.007];
	\end{tikzpicture}
	\hspace{-2mm}
}

\foreach \x in {A, ..., Z}{%
	\expandafter\xdef\csname orcid\x\endcsname{\noexpand\href{https://orcid.org/\csname orcidauthor\x\endcsname}{\noexpand\orcidicon}}
}

\begin{document}

\title[{B}ayesian wavelet de-noising]{Bayesian wavelet de-noising with the caravan prior}

\author[S.~Gugushvili]{Shota Gugushvili$^1$\orcidA{}}
\address{$^1$Biometris\\
	Wageningen University \& Research\\
	Postbus 16\\
	6700 AA Wageningen\\
	The Netherlands}
\email{shota@yesdatasolutions.com}
\thanks{The research leading to the results in this paper has received funding from the European Research Council under ERC Grant Agreement 320637.}

\author[F. H.~van der Meulen]{Frank van der Meulen$^2$\orcidB{}}
\address{$^2$Delft Institute of Applied Mathematics\\
Faculty of Electrical Engineering, Mathematics and Computer Science\\
Delft University of Technology\\
Van Mourik Broekmanweg 6  \\
2628 XE Delft\\
The Netherlands}
\email{f.h.vandermeulen@tudelft.nl}

\author[M.~Schauer]{Moritz Schauer$^3$\orcidC{}}
\address{$^3$Department of Mathematical Sciences,
	Chalmers University of Technology and University of Gothenburg,
	SE-412 96 G\"{o}teborg,
	Sweden}
\email{smoritz@chalmers.se}

\author[P.~Spreij]{Peter Spreij$^4$\orcidD{}} 
\address{$^4$Korteweg-de Vries Institute for Mathematics\\
University of Amsterdam\\
P.O. Box 94248\\
1090 GE Amsterdam\\
The Netherlands \and Institute for Mathematics, Astrophysics and Particle Physics\\ Radboud University\\ Nijmegen\\ The Netherlands}
\email{spreij@uva.nl}



\begin{abstract}

According to both domain expert knowledge and empirical evidence, wavelet coefficients of real signals tend to exhibit clustering patterns, in that they contain connected regions of coefficients of similar magnitude (large or small). A wavelet de-noising approach that takes into account such a feature of the signal may in practice outperform other, more vanilla methods, both in terms of the estimation error and visual appearance of the estimates. Motivated by this observation, we present a Bayesian approach to wavelet de-noising, where dependencies between neighbouring wavelet coefficients are a priori modelled via a Markov chain-based prior, that we term the caravan prior. Posterior computations in our method are performed via the Gibbs sampler. Using representative synthetic and real data examples, we conduct a detailed comparison of our approach with a benchmark empirical Bayes de-noising method (due to Johnstone and Silverman). We show that the caravan prior fares well and is therefore a useful addition to the wavelet de-noising toolbox. 

\end{abstract}


\maketitle
\section{Introduction}
\sloppy 
\subsection{Setup}

Let $f$ be an unknown function observed on a regularly spaced grid of $N=2^J$ points $\{t_i\}$ in the regression model
\begin{equation}
\label{eq:reg}
X_i = f(t_i) + \epsilon_i,
\end{equation}
where $\epsilon_i \iid N(0,\sigma^2)$, and the noise level $\sigma^2$ is unknown. A popular approach to inference in this model relies on an application of the Discrete Wavelet Transform (DWT) to the data $\{X_i\}$, resulting in the normal means model
\begin{equation}
\label{eq:model}
Y_{j,k}=\beta_{j,k}+\varepsilon_{j,k},
\end{equation}
where $\{Y_{j,k}\}$ are the empirical wavelet coefficients, $\{\beta_{j,k}\}$ is the parameter vector of interest formed of the wavelet coefficients of $\{f(t_i)\}$, and $\varepsilon_{j,k} \iid N(0,\sigma^2)$ are unobservable stochastic disturbances (we provide more details in Section~\ref{sec:methodology}). The observations $\{Y_{j,k}\}$ are then de-noised using one of the many possible techniques, yielding upon inversion of the wavelet transform the estimates $\{ \hat{f}(t_i) \}$ of $\{f(t_i)\}$.

A rationale for a wavelet approach to regression consists in the following (see, e.g., \cite{donoho94}): DWT typically `sparsifies' the signal $\{f(t_i)\}$, in that many wavelet coefficients $\beta_{j,k}$'s are zero, or nearly so. Since the wavelet decomposition preserves the $L^2$-norm of the signal (\cite{percival00}, equation (95d)), this implies that the transformed signal $\{\beta_{j,k}\}$ will contain some large coefficients, and a contrast with small coefficients will typically be sharper than in the original signal $\{f(t_i)\}$  (cf.~\cite{percival00}, Section 10.1). On the other hand, due to the orthogonality property of DWT, the noise $\{\epsilon_i\}$ in the original observations $\{X_i\}$ gets spread out `uniformly' in the transformed observations $\{Y_{j,k}\}$, in that one still has $\varepsilon_{j,k} \iid N(0,\sigma^2)$. Hence a small absolute magnitude of an observation $Y_{j,k}$ is likely to be an indicator of the fact that the corresponding $\beta_{j,k}$ is zero (exactly, or nearly), whereas a large value of $Y_{j,k}$ likely means that it predominantly consists of the signal $\beta_{j,k}$. This forms the basis of various  wavelet thresholding or shrinkage methods, that produce estimates of $\beta_{j,k}$'s by thresholding or shrinking small $Y_{j,k}$'s to zero as containing pure noise, and keeping large $Y_{j,k}$'s (exactly or largely) unchanged (\cite{percival00}, Section 10.2). A wavelet-based approach to non-parametric regression leads to excellent practical results due to spatial adaptation properties of wavelets (see \cite{donoho94}). However, there are situations when other	 estimators are preferable. This can happen for signals that are better representable in bases other than the wavelet basis, e.g., `frequency domain' signals such as the sinusoid. %

\subsection{Related work}

Within the Bayesian paradigm, the notion of sparsity can be naturally modelled through imposing a sparsity-inducing prior distribution on the coefficients $\{\beta_{j,k}\}$. There are two main possibilities to that end. The first is based on discrete mixtures, that model the signal $\{\beta_{j,k}\}$ via a combination of a point mass at zero and an absolutely continuous component elsewhere. The corresponding prior is often referred to as the spike-and-slab prior (see, e.g., \cite{mitchell88}). In the second approach, absolutely continuous shrinkage priors are used instead; these put a mass around zero and also exhibit heavy tails (see, e.g., \cite{tipping01} or \cite{carvalho10}). While the former approach leads to a correct representation of sparse estimation problems by placing a point mass at zero, truly sparse solutions are not possible with the latter; in case they are desired, they require a further device, e.g.\ some form of thresholding. Nevertheless, with shrinkage priors the point estimates of zero coefficients are still strongly shrunk to zero. Also, shrinkage priors are attractive computationally and have been demonstrated to perform well in various circumstances. Whether the real life signals are truly sparse in the strict sense that their small wavelet coefficients are exactly equal to zero, might be debatable.

Several Bayesian approaches to wavelet de-noising are discussed in \cite{percival00}, pp.~412--415 and 426--428. However, the method that gained the greatest acclaim in the wavelet de-noising context is the empirical Bayes method of \cite{johnstone05}, which we will refer to as EBayes. EBayes relies on the spike-and-slab prior, with its hyperparameters optimised by maximising the marginal likelihood, see \cite{silverman04}. The simulation studies in \cite{johnstone05} demonstrate overall excellent performance of EBayes, and Bayesian point estimates resulting from it possess a natural shrinkage property.   In fact, the coefficients $\beta_{j,k}$'s can even be estimated exactly as zero if the posterior median is used as a point estimate, and in that case the solution to the estimation problem is truly sparse. We thus consider EBayes as a benchmark in this article. This is in line with earlier works in the sparse normal means model, see, e.g., \cite{carvalho10} and \cite{polson11}, who studied the horseshoe prior.

\subsection{Structured sparsity}
\label{sec:setup}

\begin{figure}
\begin{center}
\includegraphics[width=0.85\textwidth]{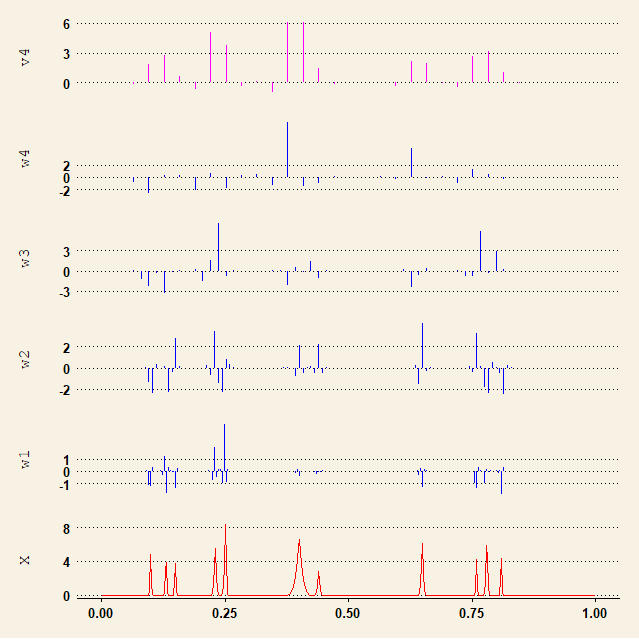}
\end{center}
\captionsetup{width=0.85\textwidth, font=small}
\caption{
DWT coefficients of $N=512$ values of the Bumps function arranged by levels of the transform. Periodic boundary conditions and the $\operatorname{LA}(8)$ filter corresponding to Daubechies' least asymmetric wavelet with $4$ vanishing moments (Symmlet $4$) are used to compute DWT. The number of computed levels of the transform is $J_0=4$. The scaling coefficients at level $4$ are displayed at the top (and can be ignored at present), followed by wavelet coefficients (from levels $4$ to $1$) and the original data. In each level, the coefficients are aligned via circular shifting so as to correspond to the events in the original signal (precise description of their arrangement is given in Sections 4.8 and 4.11 in \cite{percival00}); furthermore, the heights of vertical lines emanating from a horizontal zero line give relative sizes of coefficients, with zero coefficients not displayed. See Subsection \ref{sec:dwt} below for some additional details on DWT. For a better visibility and economy of space, the  individual panels are made of the same size, so that their vertical scales are in fact different.}
\label{fig:bumps}
\end{figure}

It has been observed in the literature that with DWT the sparsification of the signal $\{f(t_i)\}$ occurs in a structured manner. By this we mean that non-zero wavelet coefficients tend to cluster instead of being scattered in a completely random fashion across the signal $\{\beta_{j,k}\}$; see, e.g., Section 10.8 in \cite{percival00}, or Appendix \ref{sec:literature}, where we have collected several relevant quotes from the literature. Here we illustrate the phenomenon on a simple but representative example (cf.~\cite{cai01}). Consider Figure \ref{fig:bumps}, where we plotted the wavelet coefficients computed from $N=512$ values of the Bumps function (see \cite{donoho95}). It is seen from the plot that when arranged according to levels of DWT, the wavelet coefficients with large absolute magnitudes occur in clusters, namely approximately at those locations where the function undergoes abrupt changes. Additionally, many coefficients are quite small or zero.

Given that wavelet coefficients typically exhibit structures beyond `mere sparsity', it appears natural to incorporate in inferential procedures some of their additional features. A closely related question that one may ask is: Does ignoring possible local structures in the signal produce scientifically satisfactory answers? In that respect, the domain expert knowledge in, e.g., audio signal processing indicates that a failure to account for the structure of the signal in de-noising applications may result in unacceptable solutions for a human ear. Likewise, \cite{donoho95:denoise} stresses importance of reducing the extent of undesirable noise-induced structures like `ripples', `blips' and oscillations in the inferred signal, citing geophysical and astronomical studies, where such effects may lead to interpretational difficulties. Somewhat disappointingly, frequentist estimation methods that account for clustering of non-zero wavelet coefficients via block thresholding, such as NeighBlock and NeighCoeff of \cite{cai01}, have been shown to perform worse in practice than EBayes, that does not assume any additional structure beyond sparsity.

In this paper, we propose a Bayesian wavelet de-noising method that accounts for existence of special structures in wavelet coefficients. We compare it to EBayes and show via simulation and real data examples that our estimator, that we baptised the caravan estimator, measures up well to EBayes, often it  substantially as far as estimation accuracy is concerned (in terms of the square estimation error). Nevertheless, the caravan estimator does not achieve a uniform  improvement (i.e.\ over all simulation scenarios) upon EBayes.

\subsection{Organisation}

The rest of the paper is organised as follows: in Section \ref{sec:methodology} we introduce in detail the statistical problem and our Bayesian methodology to tackle it. Section \ref{sec:simulations} studies the performance of our method on synthetic data examples and compares it to the main alternative: EBayes. Section \ref{sec:nmr} deals with real data examples. Section \ref{sec:conclusions} summarises our findings and outlines directions for future research. In Appendix~\ref{sec:literature} a small compendium of quotes from the literature, illustrating some of the points we made in this paper, is presented. Appendix \ref{app:gibbs} gives details of the Gibbs sampler we use to evaluate the posterior, while Appendices~\ref{app:hyper} through \ref{app:addsim} contain further details on our simulation study. 

\subsection{Notation}
$N(\mu,\sigma^2)$ denotes the normal distribution with mean $\mu\in\mathbb{R}$ and variance $\sigma^2>0$. $\Exp(\lambda)$ is the exponential distribution with rate parameter $\lambda>0$, whose density is $x \mapsto \lambda e^{-\lambda x},$ for $x>0$. $\gam(a,b)$ is the gamma distribution with shape parameter $a$ and rate parameter $b>0$, whose density is
\[
x \mapsto \frac{b^a}{\Gamma(a)} x^{a-1}e^{-b x}, \quad x>0,
\]
where $\Gamma$ is the gamma function. The inverse gamma distribution with shape parameter $a>0$ and scale parameter $b>0$ is denoted by $\ig(a,b)$. Its density is
\[
x \mapsto \frac{b^a}{\Gamma(a)} x^{-a-1}e^{-b / x}, \quad x>0.
\]
In conformance with standard Bayesian notation, we often denote random variables with lowercase letters, such as $x$, and write the corresponding density as $p(x)$. Conditioning of $x$ on $y$ is denoted by $x\mid y$, with $p(x\mid y)$ standing for the conditional density of $x$ given $y$.

\section{Methodology}
\label{sec:methodology}

In this section we provide a detailed description of our Bayesian methodology for wavelet de-noising.

\subsection{Discrete wavelet transform}
\label{sec:dwt}

DWT is an orthogonal transformation applied on a finite dyadic sequence of numbers (that the data length $N$ is a dyadic number, $N=2^J$, say, is a restriction, although there are some ad hoc ways to deal with it; see, e.g., pp.~141--145 in \cite{percival00}). Starting with the data $x=(x_0,\ldots,x_{N-1})$, DWT can be conveniently described through successive applications of special low- and high-pass filters $\mathcal{H}=\{h_k\}$ and $\mathcal{G}=\{g_k\}$ (referred to as quadrature mirror filters) in combination with dyadic decimation or downsampling steps; jointly, these constitute the so-called pyramid algorithm. Care has to be exercised when computing DWT coefficients at the boundaries; we use periodic boundary conditions throughout. Define $v_0$ to be the original data $x$, and let $(\downarrow 2)$ be the downsampling operator. \cite{percival00} use odd decimation, retaining odd-indexed entries of a given sequence; thus for $y=(\ldots,y_{-2},y_{-1},y_0,y_1,y_2,\ldots)$, say, $(\downarrow 2)y=(\ldots,y_{-3},y_{-1},y_1,y_3,\ldots)$. This is a matter of convention, and the even decimation would have been an equally valid choice. The scaling coefficients at level $1$ are $v_1 = (\downarrow 2) \mathcal{H} v_0$, whereas the wavelet or detail coefficients are given by $w_1=(\downarrow 2) \mathcal{G} v_0.$ Here the notation $\mathcal{H} v_0$ stands for circular convolution of $v_0$ with $\mathcal{H}$, and similarly for $\mathcal{G}$. Then one proceeds inductively: with $v_j$ and $w_j$ being already defined, one sets $v_{j+1} = (\downarrow 2)\mathcal{H} v_j$ and $w_{j+1} = (\downarrow 2) \mathcal{G} v_j$. The process can be either brought to completion, the final processed level being $j=J$, or stopped at level $j=J_0 < J$; in this last case one talks about a partial DWT (a partial DWT does not require $N$ to be a dyadic number: it is enough to have that $N$ is an integer multiple of $2^{J_0}$). For a fixed $j$, the vectors $v_j$ and $w_j$ have length $N/2^j$, and their elements can be enumerated as $v_{j,k}$ and $w_{j,k}$, respectively, for $k=0,1,\ldots,N/2^j-1$. The scaling coefficients $v_{J_0}$ can be thought of as corresponding to a low-frequency component of the signal $x$, whereas the wavelet coefficients $w_1,\ldots,w_{J_0}$ to the high-frequency components. When stacked together, $v_{J_0}$ and $w_{J_0},\ldots,w_1$ constitute an orthogonal transform of the data $x$; the latter can be easily recovered via the inverse pyramid algorithm. Both DWT and its inverse can be evaluated efficiently in $O(N)$ multiplications. Conceptually, the wavelet detail coefficients $w_j$ can be associated with changes in $x$ at the scale $2^{j-1}$, i.e., loosely speaking, with differences of averages formed of $2^{j-1}$ successive values in $x$. On the other hand, $v_{J_0}$ is associated with changes in $x$ at scale $2^{J_0}$ and higher; in fact, if $J_0=J$, $v_{J_0}$ is a (rescaled) sample mean of $x$. 

\begin{figure}
	\begin{center}
	\includegraphics[width=0.85\textwidth]{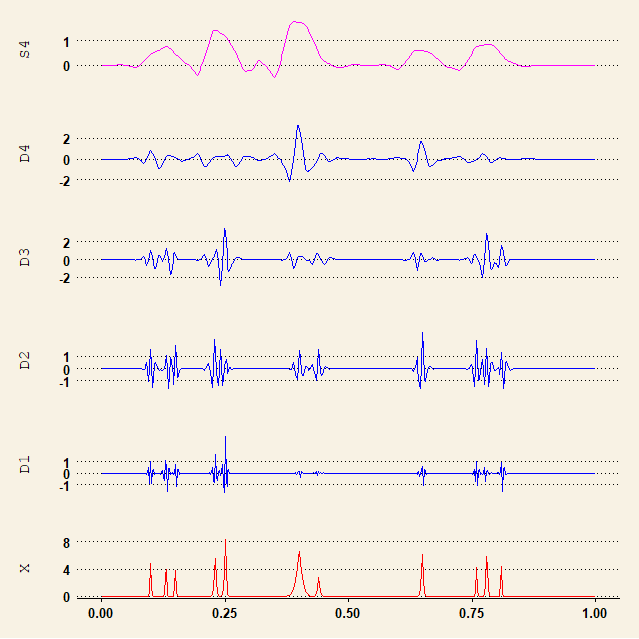}
	\end{center}
	\captionsetup{width=0.85\textwidth, font=small}
	\caption{MRA of the Bumps function discretely sampled on a uniform grid of $N=512$ points. DWT with the $\operatorname{LA}(8)$ filter and periodic boundary conditions was used, and $J_0=4$ levels of the transform were computed. The smooth  $S_4$ and details $D_j$'s are stacked on top of each other, and the bottom plot gives the original data $X$. For a better visibility, the vertical scales of the plots are made different, so that each panel is of equal height. Note that $S_4$ indeed has an appearance of a (rescaled) smooth of the data.}
	\label{fig:bumps:mra}
\end{figure}

Let $W$ be a matrix corresponding to DWT applied on data $x$. Then the vector $w = (w_1,\ldots,w_{J_0},v_{J_0})$ of wavelet and scaling coefficients can be obtained as $w = {W} x$ (analysis equation), and furthermore, due to orthogonality of $W$, $x =  {W}^T w$ (synthesis equation). It holds that
\begin{equation}
\label{eq:decomp:sum}
x = \sum_{j=1}^{J_0} {W}_j^T w_j + {V}_{J_0}^T v_{J_0} = \sum_{j=1}^{J_0} D_j + S_{J_0},
\end{equation}
where the matrices ${W}_j$, $j=1,\ldots,J_0$, and ${V}_{J_0}$ are obtained by partitioning ${W}$ into submatrices with the number of rows commensurate with $w_1,\ldots,w_{J_0},v_{J_0}$; cf.~\cite{percival00}, Sections 4.1 and 4.7. The $N$-dimensional vectors $D_j = {W}_j^T w_j$ are called wavelet details, whereas $S_{J_0} = {V}_{J_0}^T v_{J_0}$ is referred to as the $J_0$th level wavelet smooth. Together, $D_1,\ldots,D_{J_0}$ and $S_{J_0}$ define a multiresolution analysis (MRA) of $x$, which can be synthesised back from these components by a simple addition, see equation \eqref{eq:decomp:sum}. The detail $D_j$ corresponds to the portion of synthesis $x =  {W}^T w$ attributable to scale $2^{j-1}$, whereas the smooth $S_{J_0}$ can be viewed as a smoothed version of $x$ and is associated with changes at scale $2^{J_0}$ and higher. See Figure~\ref{fig:bumps:mra} for an illustration of MRA for the Bumps function.

For a detailed exposition of wavelet transforms, the reader may consult any of the numerous reference works on the topic, e.g.\ \cite{percival00}.
Furthermore, we implicitly assume that the wavelet coefficients have been realigned so as to approximately correspond to the output of a zero-phase filter. Admittedly, though, the statistical impact of the latter adjustment was not particularly noticeable in the simulation examples we considered.  A filter is called zero-phase, if its transfer function is real-valued at Fourier frequencies. This  allows to associate with $Y_i$'s the physically meaningful time scale of the original data $\{X_i\}$ (see pp.~108--110 in \cite{percival00}). For Daubechies' filters (which are the ones used in the present work), a proper alignment can be achieved by circularly shifting the output of a filtering step by a specified amount, depending on the filter and the transform level, as discussed on pp.~146--147 in \cite{percival00}. 

\subsection{Statistical model}
\label{sec:likelihood}

In our regression context, upon applying DWT on the observations $\{X_i\}$, one obtains empirical wavelet coefficients $\{Y_{j,k}\}$ arranged according to  levels $j=1,\ldots,J_0$. Recall from Equation~\eqref{eq:model} that the signal wavelet coefficients are denoted by $\{\beta_{j,k}\}$.
The statistical model for level $j$ wavelet coefficients of the data is a Gaussian sequence model,
\[
Y_i \mid \beta_i \sim N(\beta_i,\sigma^2), \quad i=1,\ldots,n,
\]
where in order to ease our notation, we have replaced the double index $j,k$ in \eqref{eq:model} with a single index $i$ (since $j$ stays fixed), and have also set $n=N/2^j$.

Following a standard wavelet de-noising approach, originally proposed in \cite{donoho94}, we will estimate the error standard deviation $\sigma$ by the median absolute deviation (MAD) computed from the finest ($j=1$) level of DWT of the data, i.e.\ the empirical wavelet coefficients $\{Y_{1,k}\}$. Intuition underlying this estimate is that the majority of wavelet coefficients of the signal $\{f(t_i)\}$ at level $1$ will be zero, so that $Y_{1,k}$'s are mostly pure noise; a few outlier non-zero entries $\beta_{1,k}$ will not affect adversely a robust estimate of the error standard deviation such as the MAD. The estimate will be denoted by $\hat{\sigma}$. In principle, upon equipping $\sigma$ with a prior, it is also possible to take a fully Bayesian approach to estimate this parameter. However, as can be seen below, our proposal is simpler, since it allows to infer our primary objects of interest, the wavelet coefficients $\{\beta_{j,k}\}$, level by level in DWT. This is convenient, e.g.\ because different levels of DWT are expected to have different sparsity degrees, or because such a subdivision of the inference problem into smaller subtasks may speed up the algorithm we propose below. Once we have estimated the wavelet coefficients, we also need the scaling coefficients at level $J_0$ in order to invert DWT and obtain an estimate of the original signal $\{f(t_i)\}$. Following \cite{donoho94}, to that end it is common to use empirical scaling coefficients computed from the data $\{X_i\}$. Thereby the portion in $\{X_i\}$ attributable to a `coarse' scale $J_0$ is automatically classified as signal (\cite{percival00}, p.~418). Estimation of scaling coefficients via empirical scaling coefficients admits a Bayesian interpretation: assuming scaling coefficients are a priori independent and equipped with a vague $N(0,\gamma)$  prior, $\gamma\rightarrow\infty$, their posteriors are again normal (conditional on the data and the error variance $\sigma^2$), with means equal to empirical scaling coefficients.

The likelihood of the data $\{Y_i\}$ in parameters $\{\beta_i\}$ (with an estimate $\hat{\sigma}$ plugged in instead of $\sigma$) is
\[
\mathcal{L}_n( \{\beta_i\} ) = (2\pi)^{-n/2} \hat{\sigma}^{-n} e^{-\sum\limits_{i=1}^n (Y_i-\beta_i)^2/(2\hat{\sigma}^2)}.
\]

\subsection{Prior}
\label{sec:prior}

\begin{figure}
\begin{center}
\includegraphics[width=0.85\textwidth]{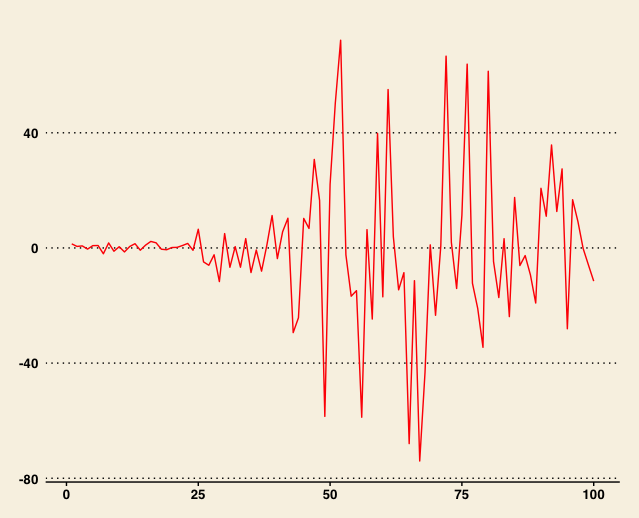}
\end{center}
\captionsetup{width=0.85\textwidth, font=small}
\caption{A realisation of the sequence $\{\beta_i\}$ of length $n=100$, using $a=5$, $\lambda_0=0.5$ and $\{\tau_i=1,i=1,\ldots,n\}$.}
\label{fig:caravan:sample}
\end{figure}

Fix hyperparameters $\{\theta_i:i=1,\ldots,n\}$, $\{\tau_i:i=1,\ldots,n\}$, and assume that a priori
\begin{equation}
\label{eq:prior:beta}
\beta_i \mid \tau_i, \theta_i \sim N\left(0, \theta_i \tau_i \right).
\end{equation}
The hyperparameters $\{ \theta_i \}$ will form an inverse gamma Markov chain, defined as follows (see \cite{cemgil07}): fix hyperparameters $a_0, b_0, a>0$, let $\{\lambda_i:i=0,\ldots,n-1\}$ be a sequence of latent variables, and consider a Markov chain
\begin{equation}
\label{eq:chain}
\lambda_0,\theta_1,\lambda_1,\theta_2,\lambda_2,\ldots,\lambda_{n-1},\theta_n
\end{equation}
with the initial and transition distributions
\begin{align*}
\lambda_0 & \sim \ig(a_0, b_0),\\
\theta_i \mid \lambda_{i-1} &\sim \ig\left( a, \frac{a}{\lambda_{i-1}} \right), \quad i=1,\ldots,n,\\
\lambda_i \mid \theta_i & \sim \ig\left( a, \frac{a}{\theta_i } \right), \quad i=1,\ldots,n-1.
\end{align*}
This definition induces a dependence structure in $\{\beta_i\}$, and ensures a degree of continuity in the absolute magnitudes of $\beta_i$'s. In fact, as explained in \cite{cemgil07}, the variables $\{\theta_i\}$ are positively correlated. Thus, e.g., a large value of $\theta_i$ is likely to go paired with a large value of $\theta_{i+1}$, which by \eqref{eq:prior:beta} increases the likelihood of a similar pairing between the absolute magnitudes of $\beta_i$ and $\beta_{i+1}$ (the latent variables $\{\lambda_i\}$ are used to achieve positive correlation between $\theta_i$'s, while retaining computational tractability of the approach; see \cite{cemgil07}). In Figure \ref{fig:caravan:sample} we display one realisation of the sequence $\{\beta_i\}$ from \eqref{eq:chain}. We do not imply that real life signals follow an inverse gamma chain, but simply that the latter provides a computationally convenient means for encoding possible dependencies present in the wavelet coefficients. The hyperparameter $a$ controls the amount of smoothing in the gamma chain, with small values corresponding to less smoothing; we assume $a \sim \gam(a_a,b_a)$. For a statistical use of inverse gamma chains outside the sparsity context see, e.g., \cite{gugu18:ppp}, \cite{gugu18:vol} and \cite{gugu18:micro}.

\begin{remark}
Note that our construction proceeds via creating dependence between absolute magnitudes of the coefficients $\{\beta_i\}$. A glance at Figure \ref{fig:bumps} shows that for stylised real-like signals, large positive coefficients may very well cluster with large negative coefficients, and in that sense our approach is natural. In fact, a similar pattern can be observed in real signals as well, such as the electrocardiogram data in Figure 127 in \cite{percival00}, but there it would have been a stretch of imagination to pretend the observations are noise-free.
\end{remark}

The parameters $\{\tau_i\}$ are local shrinkage parameters: each $\tau_i$ acts individually on $\beta_i$, and a small value of $\tau_i$ encourages shrinkage of $\beta_i$ towards zero. 
A different perspective is that this entails modelling the scale parameters with a $t$-distribution which has heavier tails than the normal distribution.
By linking $\{\tau_i\}$ via  a global shrinkage parameter $\tau_{gl}$, we introduce a global control on the sparsity level of the sequence $\{\beta_i\}$. Specifically, we assume
\[
\tau_i \mid \tau_{gl} \iid \ig(\tau_{gl},\tau_{gl}), \quad i=1,\ldots,n,
\]
with $\{\tau_i\}$ conditionally independent of other parameters in the model, given $\tau_{gl}$. In turn, the hyperparameter $\tau_{gl}$ is equipped with an independent $\gam(a_{gl},b_{gl})$ prior.

\begin{figure}
\begin{center}
\includegraphics[width=0.85\textwidth]{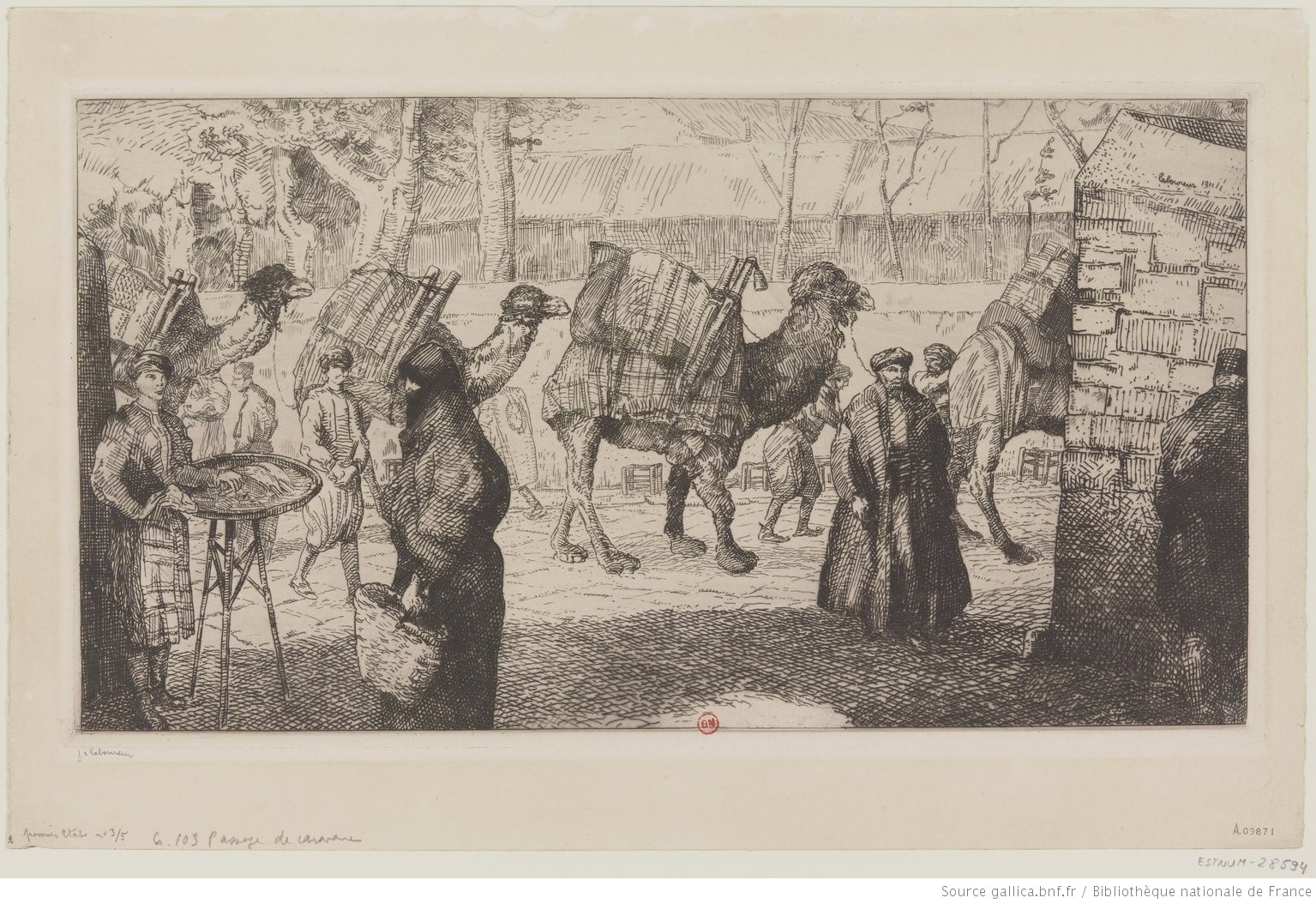}
\end{center}
\captionsetup{width=0.85\textwidth, font=small}
\caption{Passage de caravane \`a Smyrne, by Jean-\'Emile Laboureur, 1911--1912. Biblioth\`eque nationale de France, d\'epartement Estampes et photographie, FOL-EF-465 (3). Source: \url{http://gallica.bnf.fr} / BnF. Public domain.}
\label{fig:caravan}
\end{figure}

By the Markov property and the various independence assumptions we made, the joint prior on $\{\beta_i\}$, $\{\lambda_i\}$, $\{\theta_i\}$, $\{\tau_i\}$,  $\tau_{gl}$ and $a$
factorises as
\begin{multline*}
p(\tau_{gl}) \left\{ \prod_{i=1}^n p(\tau_i \mid \tau_{gl} ) \right\} \left\{ \prod_{i=1}^n p(\beta_i \mid \theta_i, \tau_i) \right\} \\
\times p(\lambda_0) p(a) \left\{ \prod_{i=1}^{n-1} p(\theta_i \mid \lambda_{i-1}, a) p(\lambda_i \mid \theta_i, a) \right\} p(\theta_n \mid \lambda_{n-1}, a).
\end{multline*}
Given the sequential nature of the definition of our prior, we term it the caravan prior, see Figure \ref{fig:caravan} for a visualisation.

\begin{remark}
Our construction of the Markov chain prior is inspired by the inverse gamma Markov chain in \cite{cemgil07}. However, it is different from the approach there, in that we also employ local shrinkage parameters $\{\tau_i\}$ linked through the global shrinkage hyperparameter $\tau_{gl}$. The two sequences $\{\theta_i\}$ and $\{\tau_i\}$ moderate or enhance each other's effects, and in a way our approach stands halfway between \cite{cemgil07} and the more conventional Bayesian approaches to wavelet de-noising proposed in the statistical literature. The parameter $a$ of the Markov chain prior fulfils a double role: on one hand it governs strength of dependence between realisations of the coefficients $\beta_i$'s; on the other hand, it affects their absolute magnitudes. A large $a$ results in a priori strongly dependent $\beta_i$'s, but also encourages them to take large values. The parameters $\{\tau_i\}$ give an additional handle to control absolute magnitudes of $\beta_i$'s, by being decoupled from the dependence structure.

A further important difference of our work from the line of research in \cite{cemgil07} and \cite{dikmen10} consists in the fact that ours concentrates on the one-dimensional wavelet transform, whereas theirs deals with transforms relevant in audio signal processing, e.g., the modified discrete cosine transform, or the Gabor transform. We provide a detailed simulation study of our approach in Section \ref{sec:simulations}, the results and conclusions of which cannot be directly read off \cite{cemgil07} and \cite{dikmen10}. Importantly, we benchmark de-noising results against the EBayes method.
\end{remark}

\begin{remark}
The idea of postulating an a priori dependence between coefficients $\{\beta_i\}$ of a sparse signal has already appeared in the statistical literature. Thus, e.g., in the audio signal processing context, \cite{wolfe04} model their parameters $\{\beta_i\}$ with the spike-and-slab prior
\[
p(\beta_i \mid \sigma_{\beta_i}, \gamma_i) = (1-\gamma_i) \delta_0(\beta_i) + \gamma_i \phi( \beta_i ; 0, \sigma_{\beta_i}^2), \quad \gamma_i \in \{0,1\},
\]
and impose a Markovian structure on the binary sequence $\{\gamma_i\}$; independent inverse gamma priors are assigned to the variances $\{\sigma_{\beta_i}^2\}$. This is different from our approach inasmuch as the spike-and-slab prior is different from the shrinkage prior.

We also mention the fact that there is a substantial body of the signal and image processing and compression literature, where dependence among wavelet coefficients is exploited in some way. See, e.g., \cite{crouse98} and references therein (this paper a priori models wavelet coefficients as discrete mixtures with a hidden state variable, and assumes the hidden states form a Markov chain).
\end{remark}

\subsection{Gibbs sampler}

The posterior for our approach is obtained from the likelihood in Subsection \ref{sec:likelihood} and the prior in Subsection \ref{sec:prior}. The posterior inference can be performed via the Gibbs sampler. In fact, as stated in Lemma \ref{lem:full:cond} in Appendix \ref{app:gibbs}, all the full conditional distributions in our model, except those of the shrinkage parameters $\tau_{gl}$ and $a$, belong to standard unimodal families and are easy to sample from. The parameters $\tau_{gl}$ and $a$ can be sampled using Metropolis-within-Gibbs steps, as explained in Appendix \ref{app:gibbs}.
Further details on this algorithm can be found, e.g., in
\cite{gelfand90}.

\section{Synthetic data examples}
\label{sec:simulations}

In this section we investigate performance of the caravan prior via representative simulation examples. Results for the DWT and MODWT de-noising are given in Subsections \ref{subsec:dwt} and \ref{subsec:modwt}. Furthermore, for readability purposes, some additional details and simulation results are deferred to Appendix \ref{app:addsim}.

\subsection{Generalities}

We implemented the caravan method in {\bf Julia} (see \cite{bezanson17}). The code is available under \cite{zenodocaravan18}.
For wavelet transforms we used the {\bf wavelets} package in {\bf R}, see \cite{wavelets} (at the moment of writing this paper, the native {\bf Julia} package for the wavelet transform is still under development), while the plots were produced with the {\bf ggplot2} package, see \cite{ggplot2}. Simulations were performed on a Macbook Air with $1.8$ GHz Intel Core i5 processor and $4$ GB $1600$ MHz DDR3 memory, running macOS High Sierra (version $10.13.5$), and on a Lenovo with $1.7$ GHz Intel Core i5-8350U processor and $8$ GB RAM, running Windows $10$ Enterprise.

Given its excellent behaviour and overall superiority over various competitors, EBayes was employed for benchmarking the caravan estimator. In short, EBayes a priori postulates that the coefficients $\beta_i \iid (1-\lambda) \delta_0(\beta_i) + \lambda p(\beta_i),$ 
where $p$ is a heavy tailed density. A Laplace density with scale parameter $a$ compares well to other possible choices of $p$. The method proceeds by estimating hyperparameters, here $\lambda$ and  $a$, by maximising the marginal likelihood, and then computing empirical Bayes estimates of $\beta_i$ (using the estimated hyperparameters). This constitutes a straightforward and numerically stable procedure.

EBayes is implemented in the {\bf EbayesThresh} package in {\bf R}, see  \cite{JS05}. We used it with settings similar to those in  \cite{JS05} and \cite{johnstone05}; in particular, an absolutely continuous part of the spike-and-slab prior assigned to wavelet coefficients $\{\beta_{j,k}\}$ was the Laplace prior with a scale parameter estimated by the empirical Bayes method, and the posterior mean and median were employed as point estimates. The wavelet transform fed to EBayes was computed via the  {\bf waveslim} package, see \cite{waveslim} (DWT computed by both the {\bf wavelets} and {\bf waveslim} packages is identical, since both packages rely on the algorithms in \cite{percival00}. However, {\bf EbayesThresh} does not support the {\bf wavelets} package; on the other hand, the latter has some functionalities we found useful). Point estimates for the caravan method were the posterior mean and median. Markov chains for the caravan method were ran for $30\, 000$ iterations ($100\, 000$ iterations for the Blocks and HeaviSine signals, see below), with the first third of the samples discarded as a burn-in. No thinning was used, but this is of course a possibility. The Metropolis-within-Gibbs steps of the caravan method were scaled to ensure acceptance rates in the range of $25-55\%$. Hyperparameters used for the caravan prior are given in Appendix \ref{app:hyper}.

Our strategy for generating noisy signals was: Sample a given function $f$ on a uniform dyadic grid of $N=512$ points $\{t_i = i/512:i=1,\ldots,512\}$, and add i.i.d.\ $N(0,\sigma^2)$ noise to the resulting values. Next, DWT was performed on the noisy data to yield the model \eqref{eq:model}. The noise standard deviation was set to $\sigma=\operatorname{SD}(\{f(t_i)\})/\operatorname{SNR}$, with $\operatorname{SD}$ standing for the sample standard deviation. We used two values for the signal-to-noise ratio: low $\operatorname{SNR}=3$  and high $\operatorname{SNR}=7$. Finally, for DWT we used the $\operatorname{LA}(8)$ filter; this choice is often reasonable in practice, see p.~136 in \cite{percival00}. The number of levels of the DWT decomposition was $J_0=6$. The quality of estimation results with DWT in fact depends on an appropriate choice of the filter, as well as the number of de-noised levels of the transform; some practical guidelines for such choices are given in Section 4.11 in \cite{percival00}. A mechanical approach to choices such as these cannot be recommended.

As the criterion to assess performance of various wavelet de-noising methods, we employed the squared error
\begin{equation}
\label{eq:sqe}
\sum_{i=1}^n (\hat{f}(t_i)-f(t_i))^2,
\end{equation}
for $\hat{f}$ an estimate of $f$, that we averaged over replicate simulation runs.

\subsection{Test functions}
\label{subsec:test}

\begin{figure}
	\begin{center}
		\includegraphics[width=0.425\textwidth]{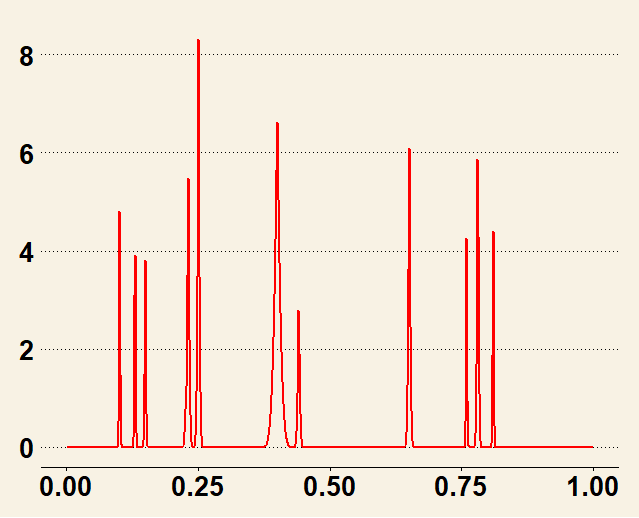}
		\includegraphics[width=0.425\textwidth]{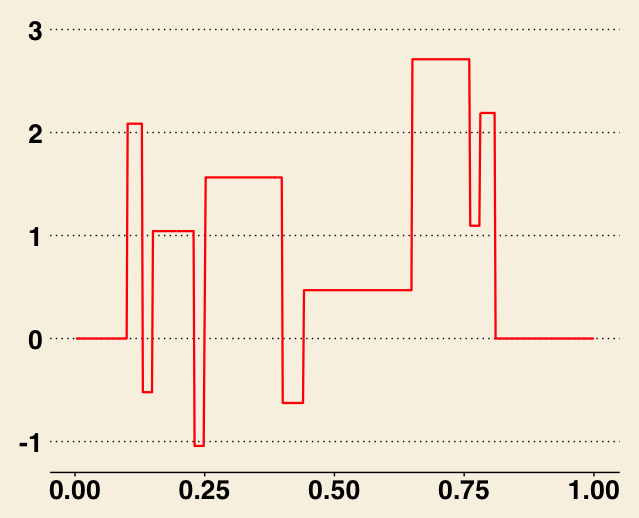}
		
		\medskip
		
		\includegraphics[width=0.425\textwidth]{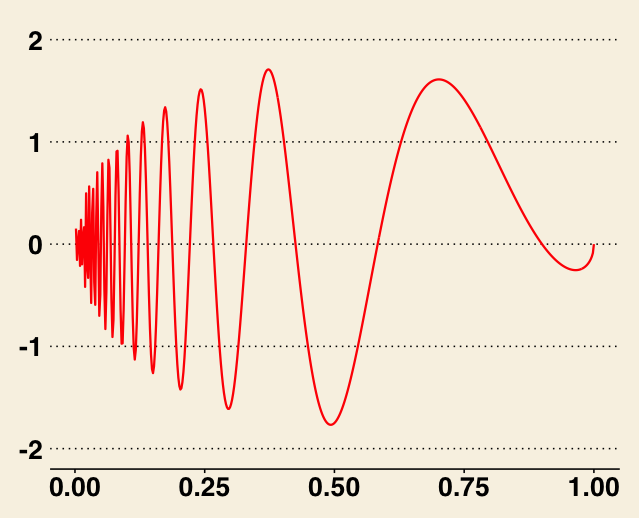}
		\includegraphics[width=0.425\textwidth]{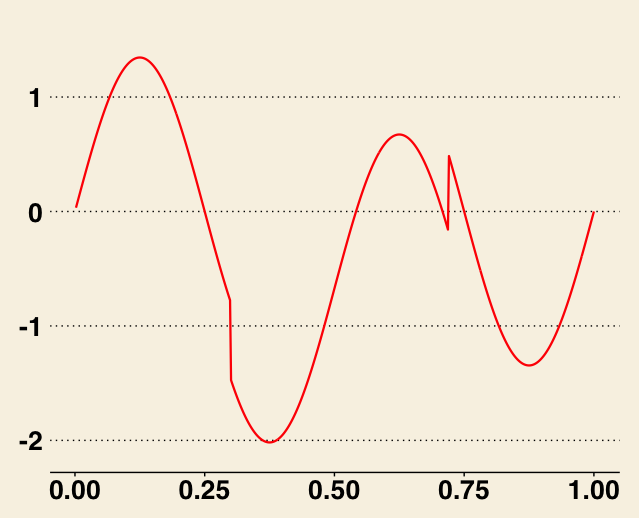}
	\end{center}
	\captionsetup{width=0.85\textwidth, font=small}
	\caption{Top row: Bumps and Blocks functions. Bottom row: Doppler and HeaviSine functions.}
	\label{fig:functions}
\end{figure}

The test functions $f$ we considered were the classical test functions named Bumps, Blocks, Doppler and HeaviSine (see \cite{donoho95}), that reproduce stylised features of signals encountered in various applications; all the expressions are collected in Appendix \ref{app:test}.
In comparison to the original definitions, we rescaled the test functions, so that the signal in each case had the standard deviation $1$. We plot the (rescaled) functions in Figure \ref{fig:functions}.

\subsection{Standard discrete wavelet transform}
\label{subsec:dwt}

We report estimation errors for the DWT (averaged over $50$ independent simulation runs) in Table1~\ref{table:dwt}, the names of the test functions there have the obvious abbreviations. While standard deviations are not displayed in these and subsequent tables, they were circa $10-20\%$ of the estimated values. It is seen from the tables that the caravan method does substantially better than EBayes for the Bumps and Doppler signals. The results are indecisive for the HeaviSine signal and equally split for the Blocks, with one of the estimators being better than another in one of the noise settings. Overall performance of the caravan method is arguably superior to that of EBayes, with the former achieving a $10-30\%$ reduction in the estimation error over the latter. Even in those cases when EBayes has a smaller estimation error, it never manages to beat the caravan estimator by too wide a margin. In terms of computational time, de-noising a single data set with the caravan method takes ca.~$1.5$ minutes (when the Gibbs sampler is run for $30\, 000$ iterations), which is reasonable on its own terms; EBayes is substantially faster, though, with its computational time being on the order of seconds instead of minutes.

{\small
	\begin{table}
		\begin{center}
			\captionsetup{width=0.85\textwidth, font=small}
			\caption{Average square errors (over $50$ simulation runs) for various test functions and methods. The sample size is $N=512$, the $\operatorname{LA}(8)$ filter is used, and periodic boundary conditions are imposed. The number of DWT levels equals $J_0=6$. The minimal average squared error in each setting is highlighted in italics and blue. The values are rounded off to one decimal after zero.}
			\begin{tabular}{lrrrr@{\hskip 0.25in}rrrr}
				\toprule 
				& \multicolumn{4}{c}{{\bf {\hskip -0.25in} Low noise}} & \multicolumn{4}{c}{{\bf High noise}}\\
				\cmidrule(l{0.1in}r{0.35in}){2-5} \cmidrule(l{0.1in}r{0.1in}){6-9}
				{\bf Method} & {\bf bmp} & {\bf blk} & {\bf dpl} & {\bf hvs}
				& {\bf bmp} & {\bf blk} & {\bf dpl} & {\bf hvs} \\
				\midrule
				Caravan (mean)   & {\color{blue} \it 3.9} & {\color{blue} \it 3.5} & { \color{blue} \it 1.8} & {\color{blue} \it 1.2} & {\color{blue} \it 21.0} & 19.4
				 & {\color{blue} \it 8.4} & {\color{blue} \it 4.0}\\
				Caravan (median) & {\color{blue} \it 3.9}  & 3.6 & {\color{blue} \it 1.8} & 1.3 & 21.3 & 20.3
				&  8.7 & 4.2\\
				EBayes (mean)    & 4.9  & 3.8 & 2.9 & {\color{blue} \it 1.2} & 22.8 & {\color{blue} \it 18.8}
				& 12.0 & 4.3\\
				EBayes (median)  & 5.6  & 4.3 &  3.3 & {\color{blue} \it 1.2}  & 25.9 & 20.6
				& 13.0 & {\color{blue} \it 4.0}\\
				\bottomrule
			\end{tabular}
			\label{table:dwt}
		\end{center}
	\end{table}
}

\begin{figure}
	\begin{center}
		\includegraphics[width=0.425\textwidth]{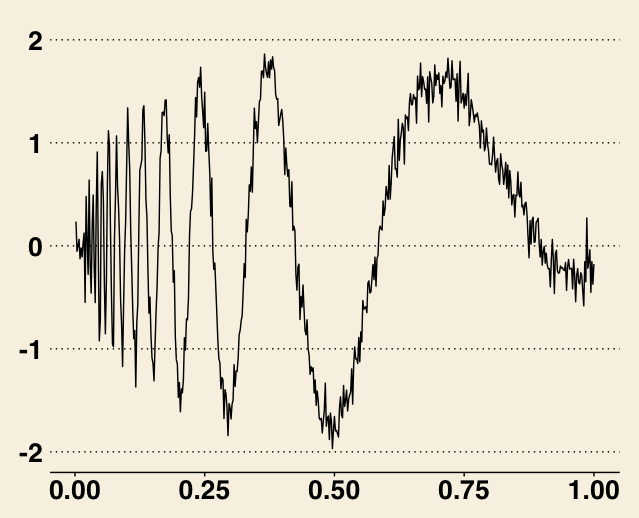}
		\includegraphics[width=0.425\textwidth]{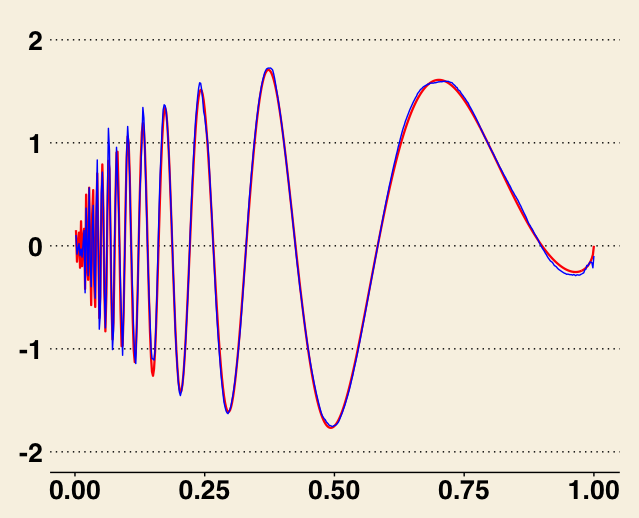}
		
		\medskip
		
		\includegraphics[width=0.425\textwidth]{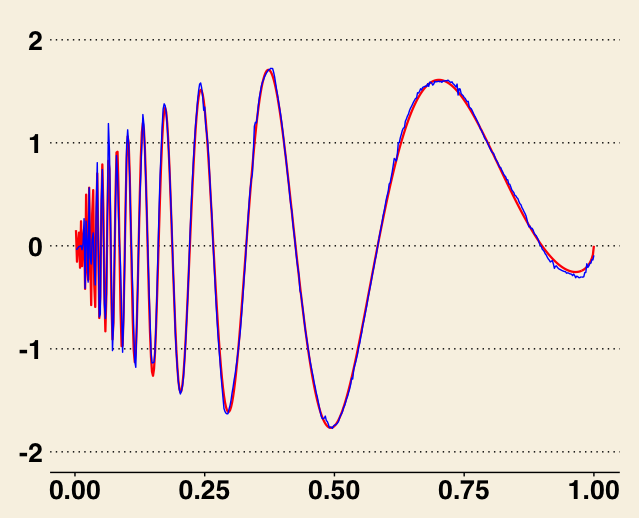}
		\includegraphics[width=0.425\textwidth]{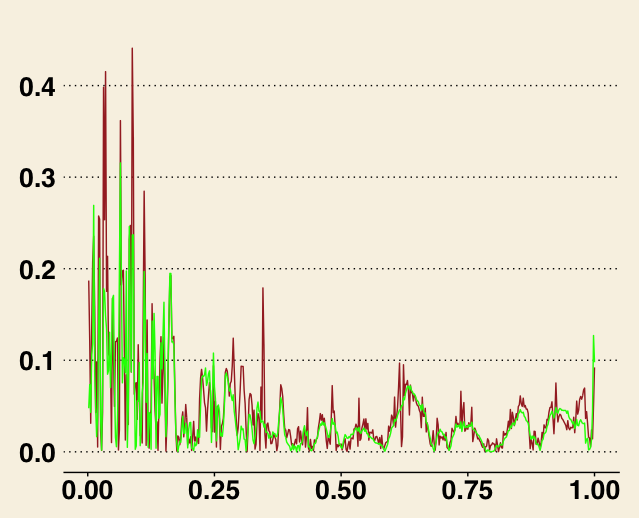}
	\end{center}
	\captionsetup{width=0.85\textwidth, font=small}
	\caption{Top row (from left to right): Noisy observations on the Doppler function (sample size $N=512$ and $\operatorname{SNR}=7$), and the caravan estimate (posterior mean) superimposed on the Doppler function. The Doppler function is in red, the estimate is in blue. Bottom row (from left to right):  EBayes (posterior mean) superimposed on the Doppler function (the colours are as in the case of the caravan estimate plot), and absolute deviations of the caravan and EBayes estimates from the Doppler function (in green and in brown, respectively). De-noising is via DWT with $J_0=6$ levels and the $\operatorname{LA}(8)$ filter.}
	\label{fig:doppler:noisy}
\end{figure}

It is instructive to display estimation results in one simulation run for the Doppler signal ($\operatorname{SNR}=7$). See Figure \ref{fig:doppler:noisy} for the noisy signal and de-noising results. The caravan estimate manages to pick up the high frequency oscillations of the signal in a neighbourhood of zero noticeably better than EBayes does. This is especially apparent from the plot of absolute deviations of both estimates from the Doppler function, and constitutes a remarkable achievement.

\begin{figure}
	\begin{center}
		\includegraphics[width=0.425\textwidth]{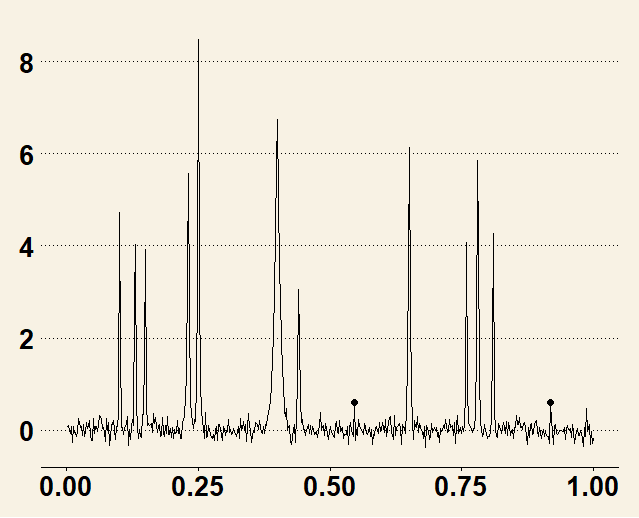}
		\includegraphics[width=0.425\textwidth]{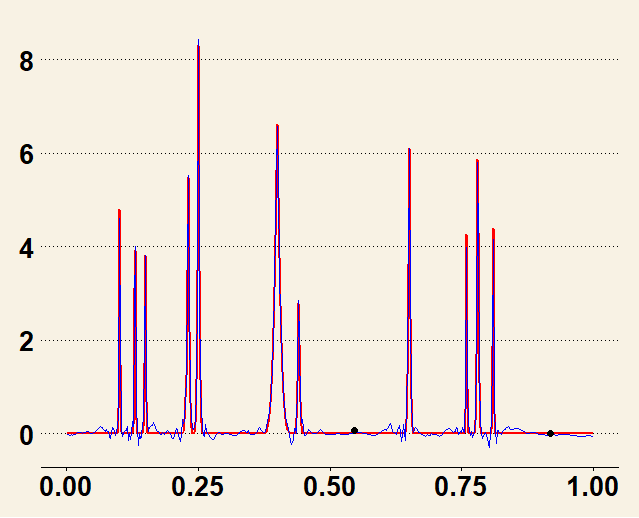}
		
		\medskip
		
		\includegraphics[width=0.425\textwidth]{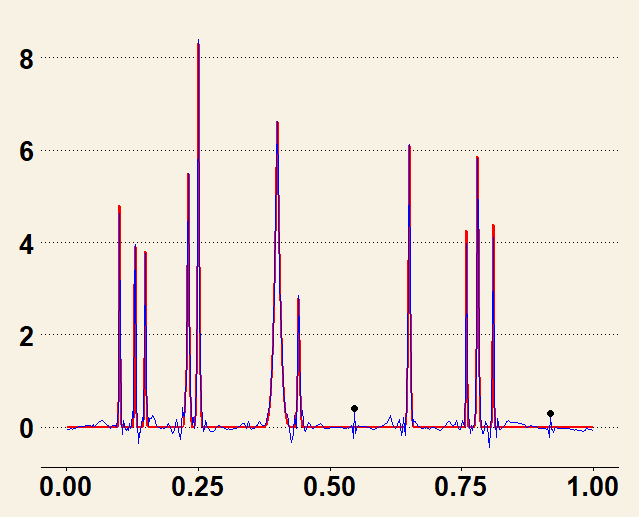}
		\includegraphics[width=0.425\textwidth]{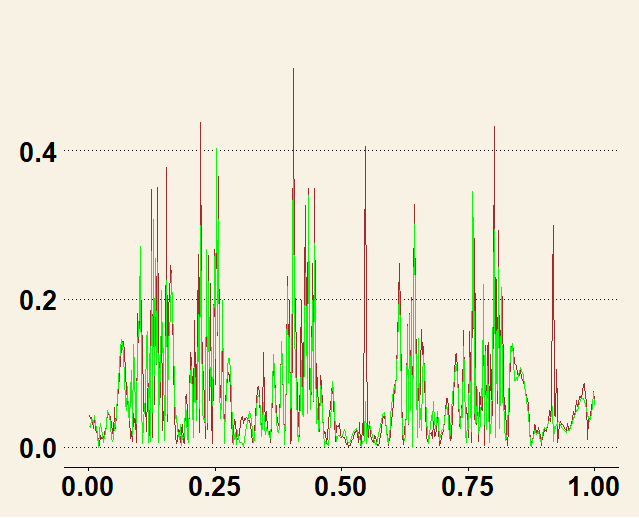}
	\end{center}
	\captionsetup{width=0.85\textwidth, font=small}
	\caption{Top row (from left to right): Noisy observations  on the Bumps function (sample size is $N=512$ and $\operatorname{SNR}=7$. The `special' points with indices $i = 280$ and $470$ that are affected by unusually large measurement errors are highlighted via large black points) and the caravan estimate (posterior mean) superimposed on the Bumps function (the true function is in red, the estimate is in blue). Bottom row (from left to right): EBayes (posterior mean) superimposed on the Bumps function (the colours are as for the caravan estimate plot), and absolute deviations of the caravan and EBayes estimates from the Bumps function (in green and in brown, respectively). De-noising is via DWT with $J_0=6$ levels and the $\operatorname{LA}(8)$ filter.}
	\label{fig:spikes1}
\end{figure}

\begin{figure}
	\begin{center}
	\includegraphics[width=0.85\textwidth]{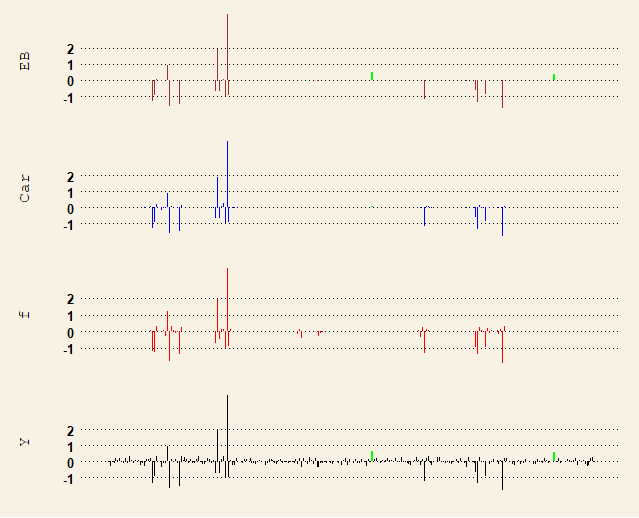}
	\end{center}
	\captionsetup{width=0.85\textwidth, font=small}
	\caption{From top to bottom: Level $j=1$ wavelet coefficients for EBayes, caravan estimate, Bumps function, and noisy observations, respectively. In each plot (excluding the one for the Bumps function itself), the pair of coefficients that is seriously affected by artificially introduced large measurement errors on top of the zero signal is highlighted in green. For details on alignment of coefficients see the caption of Figure \ref{fig:bumps}.}
	\label{fig:spikes3}
\end{figure}

To highlight one advantage of the caravan estimator over EBayes, we considered the following simulation experiment: in the $\operatorname{SNR}=7$ setting, we artificially increased measurement errors for two data points of the Bumps function in places where it is flat, in fact zero; the indices of the points were $i = 280$ and $470$.
De-noising results are reported in Figure \ref{fig:spikes1}. It is seen from the plots that among the two methods, caravan visually fares the best, in that it is the least affected by spurious peaks in the reconstructed curve due to unusually large noise on two observations. In that respect it is instructive to compare, e.g., the level $j=1$ wavelet coefficients for EBayes, caravan estimate, Bumps function, and noisy data; see Figure \ref{fig:spikes3}. As seen from that figure, two purely noise-affected empirical wavelet coefficients pass the EBayes shrinkage virtually unscathed, while they are dealt a serious blow by the caravan method.

\begin{remark}
	In relative terms, in comparison to EBayes, the Blocks and HeaviSine functions are the most difficult to de-noise with the caravan prior. Both functions are characterised by presence of discontinuities. This may be a reason for a somewhat worse performance of the caravan prior in these examples, although ascertaining a precise cause is a difficult task. In our experience, within-level dependence of wavelet coefficients, that characterises the caravan prior, appears to work less successfully when estimating the signal in a neighbourhood of a discontinuity point; conversely, in some simulation runs the caravan method was able to pick up discontinuities in a signal better than EBayes, but was then unable to perform de-noising as well as EBayes did in those regions where the signal was smooth. A better handling of signals with discontinuities via the caravan prior would require additional modelling of intra-scale dependence of wavelet coefficients. This refers to the fact that large or small values of wavelet coefficients tend to propagate across different levels of the transform, see Section~$10.8$ in \cite{percival00}; for a visualisation, see, e.g., Figure \ref{fig:bumps}. That, however, lies outside the scope of the present paper.
\end{remark}

\begin{remark}
In our experience, it is advisable to use longer Markov chain runs with the caravan prior in order to avoid visually unpleasant squiggles in de-noised curves, which in reality are solely due to the fact that the chains have not reached stationarity. Hence our decision to run the chains for $30\, 000$ or even $100\,000$ iterations (the latter is likely to be excessive in many scenarios). Giving concrete recommendations in the present context is a difficult task, as convergence of the chains depends on factors like the nature of the underlying signal, the number of observations and the signal-to-noise ratio. As one natural check, however, one can produce trace and autocorrelation plots for the hyperparameters $a,\tau_{gl}$, as well as for some of the coefficients $\beta_i$'s. See Appendix \ref{app:plots} for such plots for the Doppler signal de-noising that we considered above in Figure \ref{fig:doppler:noisy}. 

An advantage of the caravan prior is the relative simplicity of the update formulae in the Gibbs sampler (see Appendix \ref{app:gibbs}). However, this simplicity comes at a price: at each step of the sampler, only one parameter can be updated at a time, which slows down the mixing of the Markov chain for the full posterior, that is defined on a rather high-dimensional parameter space. Potentially, this may have repercussions on scalability of the method when applied on large data. See also the relevant remarks in \cite{cemgil07b} on a related Markov chain prior.
\end{remark}

\subsection{Maximal overlap discrete wavelet transform}
\label{subsec:modwt}

It has been demonstrated in, among others, \cite{coifman95}, that using the translation-invariant discrete wavelet transform for signal de-noising instead of the standard DWT often leads to better practical results, either in terms of the squared error, or visually. Unlike the standard DWT, for a data sequence of length $N$, each level of the translation-invariant transform contains $N$ wavelet coefficients, since it does not use downsampling. We specifically restrict our attention to the maximal overlap discrete wavelet transform (MODWT), see, e.g., Chapter 5 in \cite{percival00}.

\begin{figure}
	\begin{center}
		\includegraphics[width=0.85\textwidth]{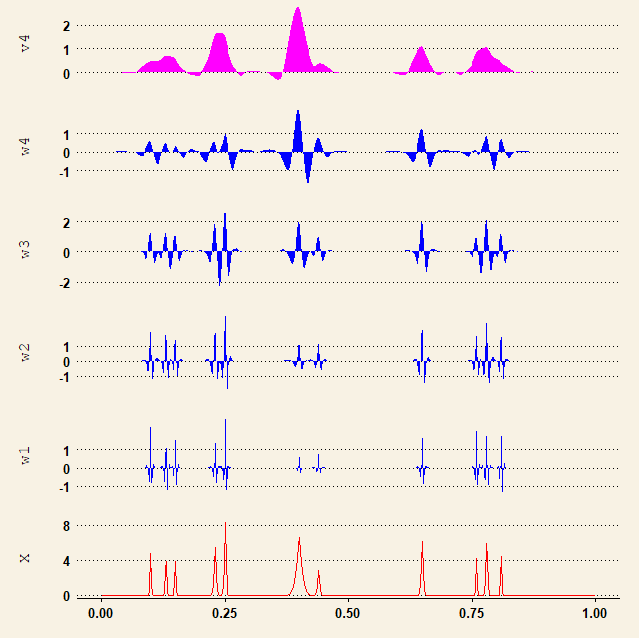}
	\end{center}
	\captionsetup{width=0.85\textwidth, font=small}
	\caption{MODWT coefficients of $N=512$ values of the Bumps function arranged by levels of the transform. The $\operatorname{LA}(8)$ filter is used. The number of computed levels of the transform is $J_0=4$, with scaling coefficients displayed at the top, and the original data at the bottom. In each level, the coefficients are aligned via circular shifting so as to correspond to the events in the original data; for precise details on the arrangement, see pp.~179--180 in \cite{percival00}.}
	\label{fig:bumps_modwt_coefficients}
\end{figure}

MODWT is highly redundant and non-orthogonal. When the data size $N$ is a dyadic number, coefficients of DWT can be extracted from those of MODWT by a suitable scaling and downsampling. Furthermore, one can extract from MODWT the coefficients of DWTs of all possible cyclic shifts of the data; see Comments and Extensions to Section 5.4 in \cite{percival00}, p.~174. Computational complexity of MODWT and its inverse (due to its redundancy, MODWT has no unique inverse; the one we have in mind is given in \cite{percival00}, and on an abstract level can be described in terms of the Moore-Penrose inverse, cf.\ p.~167 there), when evaluated via the pyramid algorithm, is $O(N\log_2 N)$ multiplications, which is somewhat slower than that for DWT, but still fast (in fact as fast as the Fast Fourier Transform). Unlike DWT, that requires the number of observations $N$ be a dyadic number, no such assumption is needed for MODWT. In theory, the number of MODWT levels $J_0$ can be arbitrarily large (unlike DWT); however, if $N$ is a dyadic integer, MODWT yields no extra information beyond the level $J=\log_2 N$, which hence can be taken as a maximal decomposition level for MODWT. See Figure \ref{fig:bumps_modwt_coefficients} for a visualisation of MODWT for the Bumps function.

Because of a lack of orthogonality, for the noisy data the MODWT wavelet coefficients will be statistically dependent. On the other hand, MODWT allows one to mitigate sensitive dependence of the standard DWT on the starting position of the data sequence (which is entirely due to downsampling used in DWT). In fact, the MODWT-based de-noising essentially performs averaging of results over all possible cyclic shifts of the data (here `all possible' means shifts by $m=0,1,\ldots,N-1$ units), that may allow a better reconstruction of the essential features of the signal and reduce noise-induced artefacts. See \cite{percival00}, Comments and Extensions to Section 5 (pp.~429--431), for a succinct description of statistical applications of MODWT.

When performing comparison of EBayes and caravan estimates, we used the settings similar to those in Section \ref{sec:simulations}. In particular, the sample size was $N=512$. We employed the $\operatorname{LA}(8)$ filter and the periodic boundary conditions. The number of levels of MODWT was $J_0=4$. Some guidelines on practicalities such as these are given in Section 5.11 in \cite{percival00}. Finally, separately for each level $j$ of MODWT, we estimated the error standard deviation $\sigma_j$ by the MAD estimate computed from the empirical wavelet coefficients of that level. It should be clear that such estimates of $\sigma_j$ cannot be expected to lead to necessarily good results in all cases, if only because the sparsity degree of MODWT (or DWT) coefficients typically decreases for coarser levels of the transform, whereas the non-zero coefficients tend to become larger (cf.~also the remarks on p.~450 in \cite{percival00}). Hence our decision to de-noise only $4$ levels of MODWT.

\begin{remark}
In the case the sample size $N$ is a dyadic number, by simple algebra that relies on the fact that DWT coefficients are rescaled and downsampled MODWT coefficients (see \cite{percival00}, equations (96d) and (169a), and page 152), an estimate of the error variance $\sigma_j^2$ can be derived as $\hat{\sigma}_j^2 = 2^{1-j} \hat{\sigma}_1^2$. Here $\hat{\sigma}_1^2$ can be obtained via MAD applied on the first level of MODWT. However, at the moment of writing this paper such an option is not envisioned for EBayes in the {\bf EBayesThresh} package, which is a primary reason why we did not employ it in our comparison.
\end{remark}

Estimation results on the same synthetic data as in Subsection \ref{sec:dwt} are reported in Table \ref{table:modwt}. A comparison with Table \ref{table:dwt} (that displayed the results for DWT) shows that MODWT substantially improves estimation accuracy of both the caravan and EBayes methods, except for the HeaviSine signal. The caravan method does better than EBayes for the Bumps, Blocks and Doppler signals. The results are indecisive for the HeaviSine function, with either method better than another in different noise settings. Overall performance of the caravan method is superior to that of EBayes, the margin being a $10-20\%$ reduction in the square error. In terms of computational time, de-noising a single data set with caravan method takes ca.~$6.5$ minutes, which is an order of magnitude slower than for EBayes.

\begin{remark}
	\label{rem:perf}
The fact that in some scenarios MODWT de-noising performs worse than DWT de-noising does not contradict earlier simulation studies in \cite{coifman95} and \cite{johnstone05}: DWT and translation-invariant DWT there differ in details from the implementations used by us (that are based on \cite{percival00}). Most importantly, we use a different error variance estimator in the MODWT case.
\end{remark}

{\small
	\begin{table}
		\begin{center}
			\captionsetup{width=0.85\textwidth, font=small}
			\caption{Average square errors (over $50$ simulation runs) for various test functions and methods. The sample size is $N=512$, the $\operatorname{LA}(8)$ filter is used, and  periodic boundary conditions are imposed. The number of MODWT levels equals $J_0=4$.}
			\begin{tabular}{lrrrr@{\hskip 0.25in}rrrr}
				\toprule 
				& \multicolumn{4}{c}{{\bf {\hskip -0.25in} Low noise}} & \multicolumn{4}{c}{{\bf High noise}}\\
				\cmidrule(l{0.1in}r{0.35in}){2-5} \cmidrule(l{0.1in}r{0.1in}){6-9}
				{\bf Method} & {\bf bmp} & {\bf blk} & {\bf dpl} & {\bf hvs}
				& {\bf bmp} & {\bf blk} & {\bf dpl} & {\bf hvs} \\
				\midrule
				Caravan (mean)		& {\color{blue} \it 3.2} & {\color{blue} \it 2.9} & {\color{blue} \it 1.5} & 1.2 & 15.6 & {\color{blue} \it 16.2}
				& 7.5 & 5.1 \\
				Caravan (median)	& {\color{blue} \it 3.2} & {\color{blue} \it 2.9} & {\color{blue} \it 1.5} & {\color{blue} \it 1.1}    & {\color{blue} \it 15.3} 	& 16.9
				& {\color{blue} \it 7.3} & 4.9 \\
				EBayes (mean)		& 3.6      & 3.0 & 2.0 & 1.2     & 17.3 	 & 17.7
				& 9.3 & 4.5 \\
				EBayes (median)    	& 3.9     & 3.2 & 2.1 & 1.2      & 18.5 	& 19.4
				& 9.5 & {\color{blue} \it 4.4}\\
				\bottomrule
			\end{tabular}
			\label{table:modwt}
		\end{center}
	\end{table}
}

\section{Nuclear magnetic resonance data}
\label{sec:nmr}

\begin{figure}
	\begin{center}
	\includegraphics[width=0.85\textwidth]{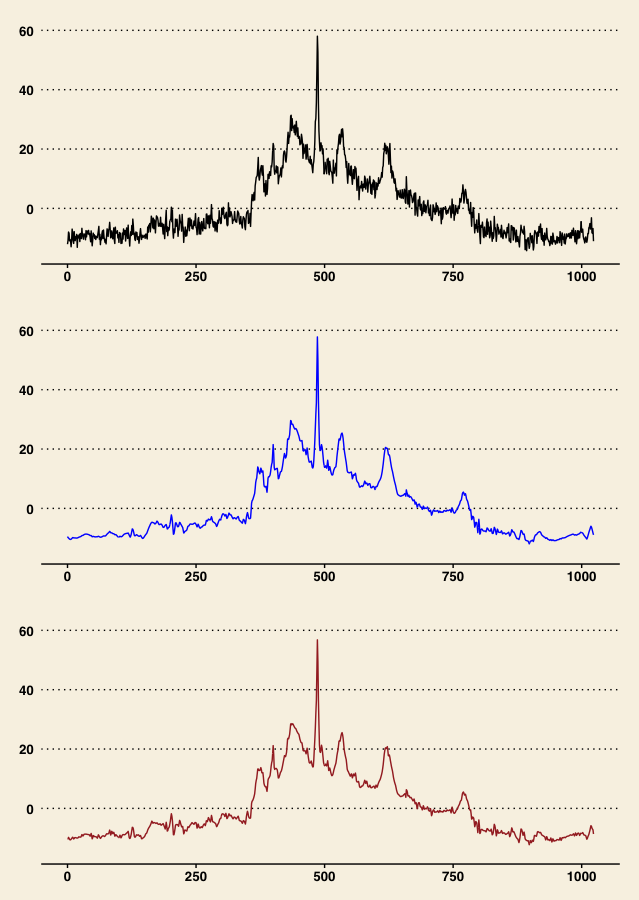}
	\end{center}
	\captionsetup{width=0.85\textwidth, font=small}
	\caption{Top panel: NMR data ($1024$ observations). Middle panel: Caravan estimate. Bottom panel: EBayes. De-noising via DWT. The $\operatorname{LA}(8)$ filter was used, with $J_0=4$ levels of the transform computed.}
	\label{fig:nmr}
\end{figure}

\begin{figure}
	\begin{center}
	\captionsetup{width=0.85\textwidth, font=small}
	\includegraphics[width=0.85\textwidth]{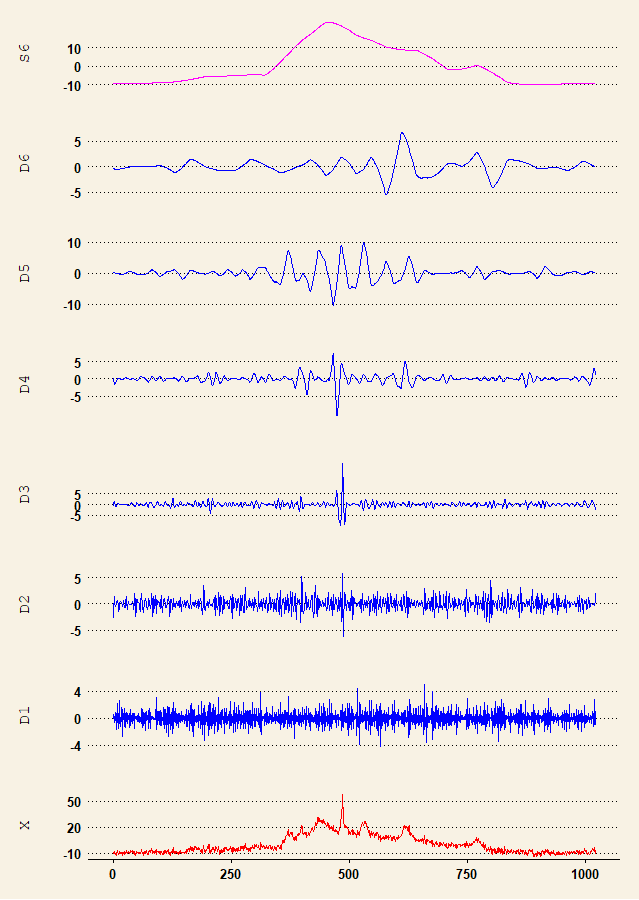}
	\end{center}
	\caption{MRA of the NMR data. DWT with $\operatorname{LA}(8)$ filter was used, and $J_0=6$ levels of the transform were computed. The top plot gives the smooth $S_6$, followed by the details $D_j$ stacked on top of each other, and the original data $X$ at the bottom.}
	\label{fig:nmr:mra}
\end{figure}

\begin{figure}
	\begin{center}
	\captionsetup{width=0.85\textwidth, font=small}
	\includegraphics[width=0.85\textwidth]{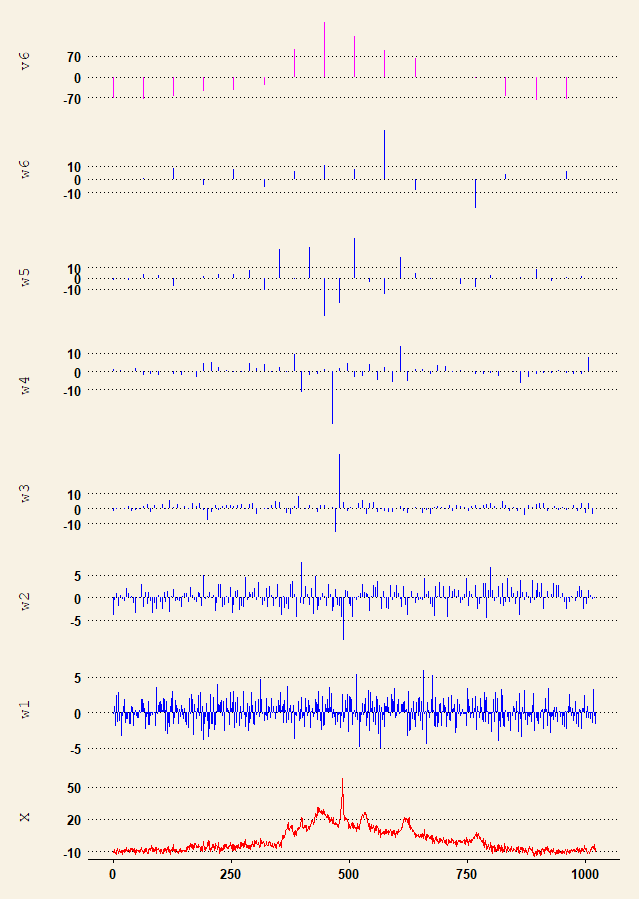}
	\end{center}
	\caption{
		DWT coefficients of the NMR data arranged by levels of the transform. Periodic boundary conditions and $\operatorname{LA}(8)$ filter are used to compute DWT. The number of computed levels of the transform is $J_0=6$. The scaling coefficients at level $6$ are displayed at the top, followed by wavelet coefficients (from levels $6$ to $1$) and the original data. See Figure \ref{fig:bumps} for additional information on the arrangement of the coefficients.}
	\label{fig:nmr:coeffs}
\end{figure}

In this section we apply our de-noising methodology on the nuclear magnetic resonance (NMR) spectrum, that constitutes a standard test data set for wavelet de-noising algorithms.\footnote{We downloaded the data from Donald B.\ Percival's website at \url{http://faculty.washington.edu/dbp/s530/} (accessed on 28 June 2018).} There are $N=1024$ observations in total, that we display in the top panel of Figure \ref{fig:nmr}. We followed Section $10.5$ in \cite{percival00}, and used the $\operatorname{LA}(8)$ filter to compute DWT. Percival and Walden de-noise $J_0=6$ levels of the transform; an MRA plot of the data set, see Figure~\ref{fig:nmr:mra}, suggests that de-noising $J_0=4$ levels of the transform might be enough. A plot of the DWT coefficients, see Figure \ref{fig:nmr:coeffs}, indicates that there are some small wavelet coefficients present at level $j=5$ too, but we opted to leave the levels $j=5,6$ as such.

In visualising de-noising results, we used posterior medians as our point estimates (we produced larger plots to clearly highlight differences between the estimates). The Markov chain for the caravan prior was run for $120\, 000$ iterations, with the first third of samples dropped as a burn-in. Both caravan and EBayes estimates remove a substantial amount of noise from the data, see Figure \ref{fig:nmr}. However, visually the caravan reconstruction appears to be more regular than EBayes. One established way to measure efficacy of a de-noising procedure in this context is to determine which of the methods better maintains the peaks of the curve; these peaks contain important information on the tissue from which the sample arose. We can compare the heights of the highest peak, cf.\ \cite{johnstone05}, p.~1719, and \cite{percival00}, p.~430. In that respect, the caravan estimate yielded the peak height $57.78$, while EBayes the peak height $56.78$. The latter method was hence worse than its competitor (to put things in perspective, the original noisy data had the peak height $58.02$).

\begin{figure}
	\begin{center}
	\captionsetup{width=0.85\textwidth, font=small}
	\includegraphics[width=0.85\textwidth]{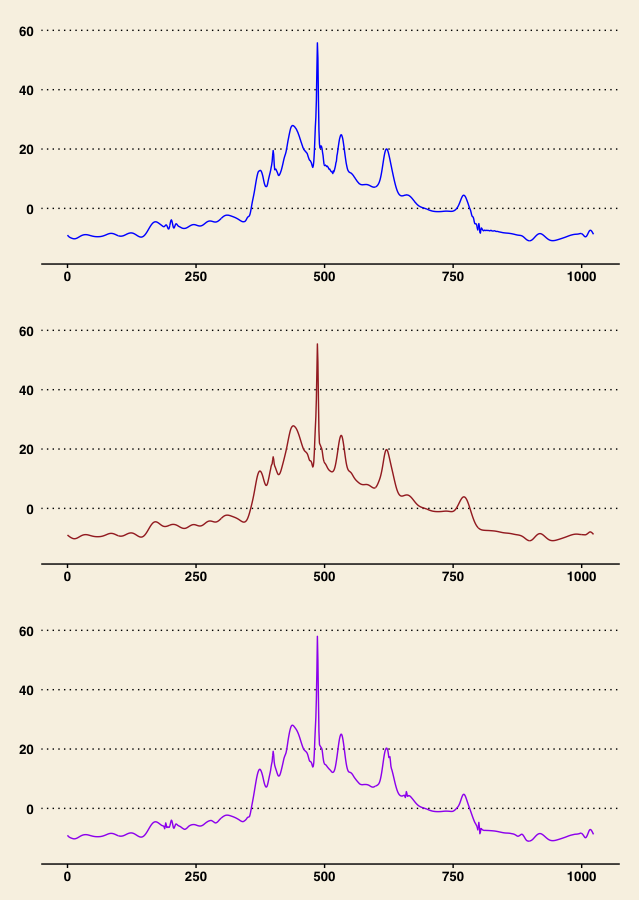}
	\end{center}
	\caption{Top panel: Caravan estimate for the NMR data. Middle panel: EBayes. Bottom panel: Hard thresholding estimate (with universal threshold). De-noising via MODWT. The $\operatorname{LA}(8)$ filter was used, with $J_0=4$ levels of the transform computed.}
	\label{fig:nmr:modwt}
\end{figure}

We also applied the MODWT de-noising (with $J_0=4$ levels), cf.~\cite{percival00}, Comments and Extensions to Section 10.5. The results are reported in Figure \ref{fig:nmr:modwt}. Both methods are even more successful in removing the noise. Concerning the highest peak, with the peak height $55.77$, the caravan estimate marginally outperformed EBayes, that yielded the peak height $55.41$. Note also how the second sharp peak to the left of the highest peak is much lower in the EBayes estimate, unlike in the caravan estimate. On the other hand, the caravan estimate shows some small squiggles near $t=200$ and $800$, that are absent in the EBayes estimate; this is similar to the hard thresholding estimate in Figure $430$ of \cite{percival00}. We reproduce that plot in the bottom panel of Figure \ref{fig:nmr:modwt}; note the appearance of an additional squiggle near $t=650$ there. Finally, a wave-like behaviour of  both estimates over the time interval $[0,300]$ is due to our decision to de-noise only $4$ levels of the transform. These waves can be largely flattened out by de-noising a $J_0=6$ level MODWT, but that would have diminished even further the heights of the sharp peaks.

Summarising, each method appears to have its own advantages on this challenging real data set.

\section{Discussion}
\label{sec:conclusions}

In this paper we studied a Bayesian approach to wavelet de-noising via a prior relying on the inverse gamma Markov chain (cf.~\cite{cemgil07}). Various types of Markov chain priors have been used for de-noising purposes in several references, but to the best of our knowledge, our paper is the first thorough comparative study of the performance of this kind of a prior. In particular, we benchmarked our method against a popular empirical Bayes procedure of \cite{johnstone05}.

Our method, which we call the caravan, strikes a good balance between conceptual simplicity and computational feasibility. Specifically, the posterior inference can be performed via a straightforward version of the Gibbs sampler. In the synthetic data examples that we considered, the method measures up well to EBayes, often substantially outperforming it in terms of the squared estimation error. The improvement brought by the caravan method comes thanks to the fact that it takes into account some of the local structures empirically observed in wavelet coefficients of real life signals. However, the caravan method does not achieve a uniform  improvement (i.e.\ over all simulation scenarios) upon EBayes, which can be taken as indication of a general excellence of the latter, rather than of a failure of the former. In particular, in our simulations the caravan prior seemed to be somewhat worse than EBayes at handling signals with jump discontinuities.

On purely visual grounds, the caravan estimator appeared to be less prone to display artefacts in its reconstructions that are due to unusually large noise peaks. As far as the computational time is concerned, since the caravan estimator is evaluated via an MCMC algorithm (Gibbs sampler), its computation is considerably slower than that of EBayes, although the method is still reasonably fast.

We believe that our paper adds a valuable Bayesian technique to the wavelet, or more generally the non-parametric regression toolbox. Furthermore, our hope is that the present contribution provides sufficient motivation for further study of the caravan method, a task that we ourselves plan to address in subsequent research. A natural question in this context, that we do not address in the present work, is: what about asymptotic statistical theory for the caravan prior? Such work in the spirit of \cite{ghosal17} has been done for the horseshoe prior in \cite{vanderpas14} and \cite{vanderpas17}. This is a problem we would very much like to study in another work.

\appendix
\section{}
\label{sec:literature}

In this appendix we present quotes from the image and signal processing and statistics literature, evidencing awareness of the need to model explicitly the structure of the signal in wavelet de-noising applications.

\begin{itemize}[leftmargin=*]
\item ``Wavelets are known for their excellent compression and localization properties. In very many cases of interest, information about a function is essentially contained in a relatively small number of large coefficients. Figure $1$ displays the wavelet coefficients of the well-known test function Bumps (Donoho and Johnstone, 1994). It shows that large coefficients come as groups; they cluster around the areas where the function changes significantly.

This example illustrates the motivation for our methods -- a coefficient is more likely to contain signal if neighbouring coefficients do also. Therefore when the observations are contaminated with noise, estimation accuracy might be improved by incorporating information on neighbouring coefficients." (\cite{cai01}).
\item ``The use of priors that can capture the dependence between the coefficients of the representation is a more delicate problem which involves \emph{expert a priori knowledge}. \ldots  Many empirical studies have concluded that the wavelet coefficients (even if the transformation is maximally decimated) of natural images are strongly dependent." (\cite{moulines04}).
\item ``It turns out that term-by-term sparsity is usually not enough to obtain state-of-the-art results both for de-noising and inverse problems involving natural images. Indeed, wavelet coefficients of images are not only sparse, they typically exhibit local dependencies among neighboring coefficients. Geometric features (edges, textures) are poorly sparsified by isotropic multiscale decompositions and create such dependencies." (\cite{peyre11}).
\end{itemize}

\section{}
\label{app:gibbs}

\subsection{Full conditionals}

\begin{lemma}
	\label{lem:full:cond}
	Define
	\[
	\xi_1 = \xi_1(\theta_i,\tau_i,\hat{\sigma}^2) =  \frac{1}{1/(\theta_i\tau_i)+1/\hat{\sigma}^2}, \quad
	\xi_2 = \xi_2(\xi_1,\hat{\sigma}^2,Y_i) = \frac{Y_i}{\hat{\sigma}^2\xi_1 }.
	\]
	The following facts hold:
	\begin{itemize}[leftmargin=*]
		\item The full conditionals for $\beta_i$, $i=1,\ldots,n$, are
		\[
		\beta_i \mid \theta_i, \tau_i,
		Y_i \sim N(\xi_2, \xi_1).
		\]
		\item The full conditionals for $\theta_i$, $i=1,\ldots,n-1$, are
		\[
		\theta_i \mid  a, \beta_i, \lambda_{i-1}, \lambda_i, \tau_i \sim \ig\left(2a + \frac{1}{2}, \frac{a}{\lambda_{i-1}} + \frac{a}{\lambda_i} + \frac{\beta_i^2}{2\tau_i} \right).
		\]
		\item The full conditional for $\theta_n$ is
		\[
		\theta_n \mid  a, \beta_n, \lambda_{n-1}, \tau_n \sim \ig\left( a + \frac{1}{2}, \frac{a}{\lambda_{n-1}} + \frac{\beta_n^2}{2\tau_n} \right).
		\]
		\item The full conditional for $\lambda_0$ is
		\[
		\lambda_0 \mid  a, \theta_1 \sim \ig\left( a_0 + a , b_0 + \frac{a}{\theta_1 } \right).
		\]
		\item The full conditionals for $\lambda_i$, $i=1,\ldots,n-1$, are
		\[
		\lambda_i \mid a, \theta_i, \theta_{i+1} \sim \ig\left( 2 a, \frac{a}{\theta_i} + \frac{a}{\theta_{i+1} } \right).
		\]
		\item The full conditional for $\tau_i$ is
		\[
		\tau_i \mid \beta_i, \theta_i, \tau_{gl} \sim \ig\left( \tau_{gl} + \frac{1}{2}, \tau_{gl} + \frac{\beta_i^2}{2\theta_i} \right).
		\]
		\item The full conditional for $\tau_{gl}$ is
		\[
		\tau_{gl} \mid \{\tau_i\} \propto \Gamma(\tau_{gl})^{ -n } \tau_{gl}^{n\tau_{gl} + a_{gl} - 1} \left( \prod_{i=1}^n \tau_i \right)^{-\tau_{gl}} \exp\left(-\tau_{gl} \left\{ b_{gl} + \sum_{i=1}^n \frac{1}{\tau_i} \right\} \right).
		\]
		Hence, up to an additive constant independent of $\tau_{gl}$, the logarithm of the full conditional is proportional to
		\[
		-n \log\Gamma( \tau_{gl} ) + ( n \tau_{gl} + a_{gl} -1 ) \log \tau_{gl} - \tau_{gl} \left\{ b_{gl} + \sum_{i=1}^n \left( \log \tau_i + \frac{1}{\tau_i} \right) \right\}.
		\]
		\item The full conditional for $a$ is
		\begin{multline*}
		a \mid \{\theta_i\}, \{\lambda_i\} \propto a^{a_a -1 + (2n-1) a} \Gamma(a)^{-(2n-1)} \prod_{i=1}^{n-1} (\theta_i^2 \lambda_{i-1} \lambda_i)^{-a} \\
		\times ( \lambda_{n-1} \theta_n )^{-a} \exp\left(-a \left\{ b_a + \sum_{i=1}^{n-1} \left( \frac{1}{\lambda_{i-1}\theta_i} + \frac{1}{\lambda_i \theta_i} \right ) + \frac{1}{\lambda_{n-1}\theta_n} \right\} \right).
		\end{multline*}
		Hence, up to an additive constant independent of $a$, the  logarithm of the full conditional is proportional to
		\begin{multline*}
		(a_a - 1 +(2n-1) a) \log a - (2n-1) \log \Gamma(a) \\
		- a \left\{ \sum_{i=1}^{n-1} \log(\theta_i^2 \lambda_{i-1}\lambda_i) + \log(\lambda_{n-1}\theta_n) 
		+ b_a + \sum_{i=1}^{n-1} \left( \frac{1}{\lambda_{i-1}\theta_i} + \frac{1}{\lambda_i \theta_i} \right ) + \frac{1}{\lambda_{n-1}\theta_n} \right\}.
		\end{multline*}
	\end{itemize}
\end{lemma}

The proof of the lemma is lengthy but elementary, and is omitted. The Gibbs sampler cycles through the above update formulae to generate approximate samples from the posterior.

\subsection{Metropolis-within-Gibbs for updating $\tau_{gl}$ and $a$}

Here we outline the Metropolis-within-Gibbs step to update the hyperparameters $\tau_{gl}$ and $a$ within the Gibbs sampler described in the previous subsection. We consider the case of $\tau_{gl}$, and note that $a$ can be treated in the same manner. Reparametrise $\tau_{gl}$ as $\widetilde{\tau}_{gl} = \log \tau_{gl}$, and observe that if $\pi$ is the full conditional of $\tau_{gl}$, the full conditional of $\widetilde{\tau}_{gl} $ is $ \widetilde{\pi} ( \widetilde{\tau}_{gl} ) = e^{ \widetilde{\tau}_{gl} } \pi ( e^{ \widetilde{\tau}_{gl} } )$. It is enough to sample $\widetilde{\tau}_{gl}$: samples ${\tau}_{gl}$ can then be obtained by simple exponentiation. Given a current value $\widetilde{\tau}_{gl}$, we propose a move
\[
\widetilde{\tau}_{gl}^{\circ} = \widetilde{\tau}_{gl} + h Z,
\]
where $h>0$ is a tuning parameter and $Z \sim N(0,1)$; this is a Gaussian random walk proposal. The acceptance probability is computed as
\[
A(\widetilde{\tau}_{gl}^{\circ} \mid \widetilde{\tau}_{gl}) = \min \left( \frac{ \widetilde{\pi} ( \widetilde{\tau}_{gl}^{\circ} ) }{ \widetilde{\pi} ( \widetilde{\tau}_{gl} ) } , 1 \right).
\]
The move is accepted, if $\log U < \log A(\widetilde{\tau}_{gl}^{\circ} \mid \widetilde{\tau}_{gl})$, where $U \sim \operatorname{Uniform}(0,1)$ is independent of $Z$; otherwise the chain stays in $\widetilde{\tau}_{gl}$. The acceptance rate is controlled by the parameter $h$.

\section{}
\label{app:hyper}

Here we give the hyperparameters for the caravan prior used in our synthetic data examples:
\[
a_a =0.1, \quad b_a = 0.1, \quad a_0 = 0.1, \quad b_0 = 0.1, \quad a_{gl} = 0.1, \quad b_{gl} = 0.1.
\]
These can be viewed as non-informative. We note that different choices may be appropriate in settings other than those considered by us. On the other hand, our specific choice appears to be quite robust, since it yielded reasonable de-noising results on different test functions, different signal-to-noise ratios, different numbers of processed levels $J_0$, different sample sizes $N$, and different individual levels of the wavelet transform.

The scaling parameters for the Metropolis-Hastings steps within our Gibbs sampler were set to $h_a = c_a/\log_2 n$ and $h_{gl} = c_{gl}/\log_2 n$, where $n$ was the number of coefficients in a given level of the wavelet transform ($h_a$ corresponds to the hyperparameter $a$, while $h_{gl}$ to the hyperparameter $\tau_{gl}$). The constants $c_a$ and $c_{gl}$ can vary per case of the underlying signal:  we used the values $c_a = 1.5$ and $c_{gl} = 2.5$.

\section{}
\label{app:test}

Here we supply definitions of the test functions that we used, upon rescaling, in our synthetic data examples.

\subsection{Bumps} The Bumps function, also introduced as a motivating example in Subsection \ref{sec:setup}, is given by
\[
f(t)= \frac{1}{7}\sum_{j=1}^{11} h_j K\left(\frac{t-t_j}{w_j}\right), \quad t\in[0,1],
\]
where $K(t)=(1+|t|)^{-4}$ and
\begin{align*}
\{t_j\}&=(0.1, \, 0.13, \, 0.15, \, 0.23, \, 0.25, \, 0.4, \, 0.44, \, 0.65, \, 0.76, \, 0.78, \, 0.81),\\
\{h_j\}&=(4, \, 5, \, 3, \, 4, \, 5, \, 4.2, \, 2.1, \, 4.3, \, 3.1, \, 5.1, \, 4.2),\\
\{w_j\}&= \{ 0.005, \, 0.005, \, 0.006, \, 0.01, \, 0.01, \, 0.03, \, 0.01, \, 0.01, \, 0.005, \, 0.008, \, 0.005 \}.
\end{align*}
Donoho and Johnstone view the Bumps function as a stylised example of a spectrum arising, e.g., in nuclear magnetic resonance (NMR) spectroscopy; see \cite{donoho95}.

\subsection{Blocks} The Blocks function is defined as
\[
f(t)=\sum_{j=1}^{11} h_j K(t-t_j), \quad t\in[0,1],
\]
where $K(t)=(1+\operatorname{sgn}(t))/2$, where 
\[
\operatorname{sgn}(t)=
\begin{cases}
+1 & \mbox{if $t>0$},
\\
-1 & \mbox{if $t<0$},
\\
0 & \mbox{if $t=0$},
\end{cases}
\]
and where the
$\{t_j\}$ are as for the Bumps function, and
\[
\{h_j\}=(4, \, -5, \, 3, \, -4, \, 5, \, -4.2, \, 2.1, \, 4.3, \, -3.1, \, 2.1, \, -4.2).
\]
According to \cite{donoho95}, the Blocks function caricatures the acoustic impedance of layered medium in geophysics, as well as one-dimensional profiles along images in certain image processing applications.

\subsection{Doppler} The Doppler function is a sinusoid with a changing amplitude and frequency,
\[
f(t)=\sqrt{t(1-t)}\sin\left(\frac{2\pi (1+0.05}{t+0.05}\right), \quad t\in[0,1].
\]

\subsection{HeaviSine} The HeaviSine function is a sinusoid with two jumps,
\[
f(t)=4\sin (\pi t) - \operatorname{sgn}(t-0.3) - \operatorname{sgn}(0.72 - t), \quad t\in[0,1].
\]

\section{}
\label{app:plots}

Here we present several trace, autocorrelation and density plots of posterior samples for de-noising the Doppler signal, that we studied in Section \ref{sec:simulations} (see in particular Figure \ref{fig:doppler:noisy} there).

\begin{figure}
	\begin{center}
		\includegraphics[width=0.85\textwidth]{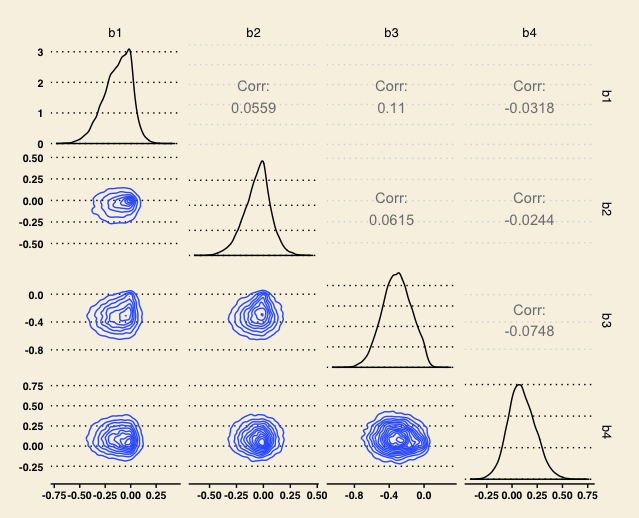}
		
		\medskip
		
		\includegraphics[width=0.85\textwidth]{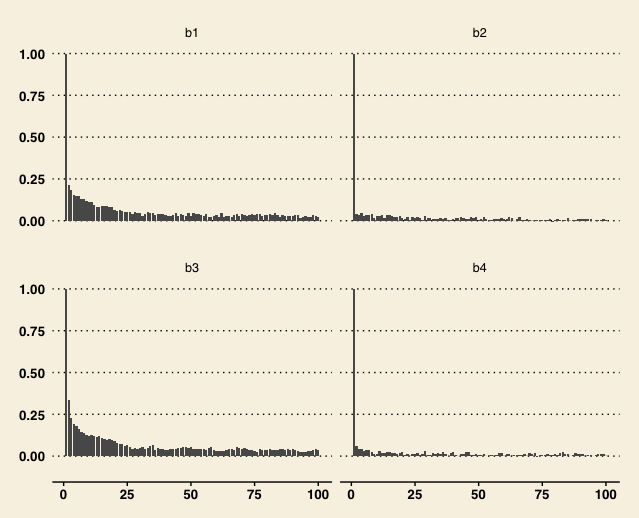}
	\end{center}
	\captionsetup{width=0.85\textwidth, font=small}
	\caption{Graphical convergence diagnostics for de-noising the Doppler function in Figure \ref{fig:doppler:noisy}. Top: Density plots and pairwise density contour plots for coefficients $\beta_1,\ldots,\beta_4$ from level $j=3$ of DWT. Numbers in the upper triangle of the plot matrix give correlations between various chains. Bottom: Autocorrelation plots.}
	\label{fig:beta:diag}
\end{figure}

\begin{figure}
	\begin{center}
		\includegraphics[width=0.85\textwidth]{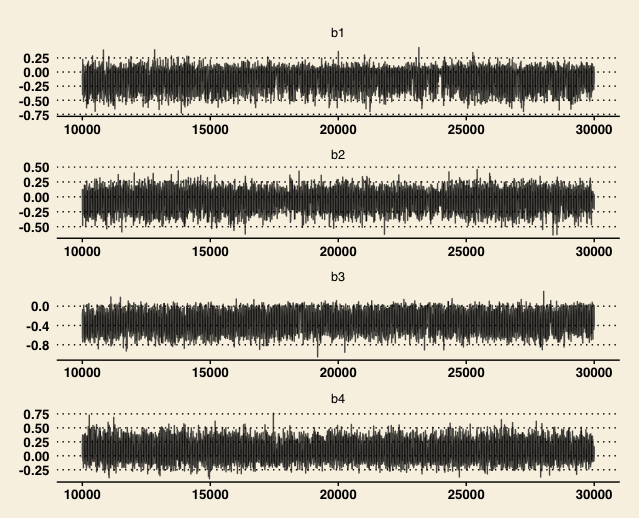}
	\end{center}
	\captionsetup{width=0.85\textwidth, font=small}
	\caption{Graphical convergence diagnostics for de-noising the Doppler function in Figure \ref{fig:doppler:noisy}. Displayed are trace plots for coefficients $\beta_1,\ldots,\beta_4$ from level $j=3$ of DWT.}
	\label{fig:beta:trace}
\end{figure}

We display these plots in Figures \ref{fig:beta:diag} and \ref{fig:beta:trace} for wavelet coefficients $\beta_1,\ldots,\beta_4$ from level $j=3$ of DWT (this choice is arbitrary). Plots were produced with the {\bf ggmcmc} package in {\bf R}, see \cite{ggmcmc}. The figures suggest that the chains are rather well-behaved and mixing.

\begin{figure}
	\begin{center}
		\includegraphics[width=0.85\textwidth]{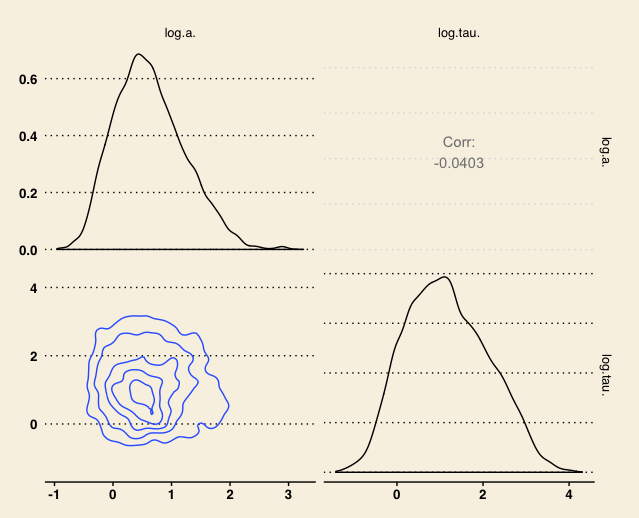}
		
		\medskip
		
		\includegraphics[width=0.85\textwidth]{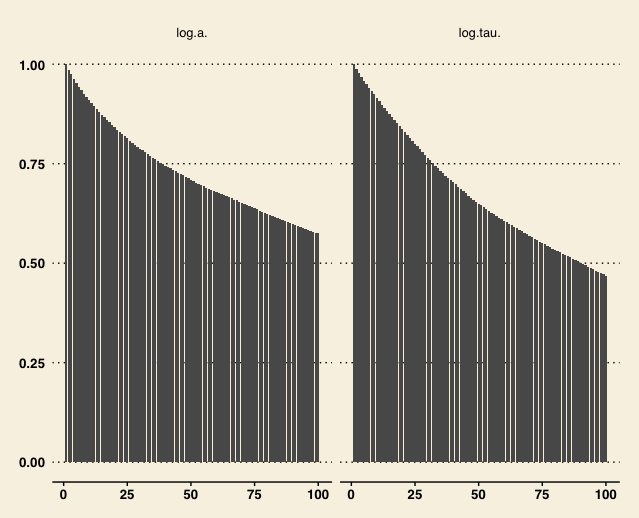}
	\end{center}
	\captionsetup{width=0.85\textwidth, font=small}
	\caption{Graphical convergence diagnostics for de-noising the Doppler function in Figure \ref{fig:doppler:noisy}. Top: Density plots and pairwise density contour plots for logarithms of hyperparameters $a$ and $\tau_{gl}$. Number in the upper triangle of the plot matrix gives correlation between the two chains. Bottom: Autocorrelation plots.}
	\label{fig:hyper:diag}
\end{figure}

\begin{figure}
	\begin{center}
		\includegraphics[width=0.85\textwidth]{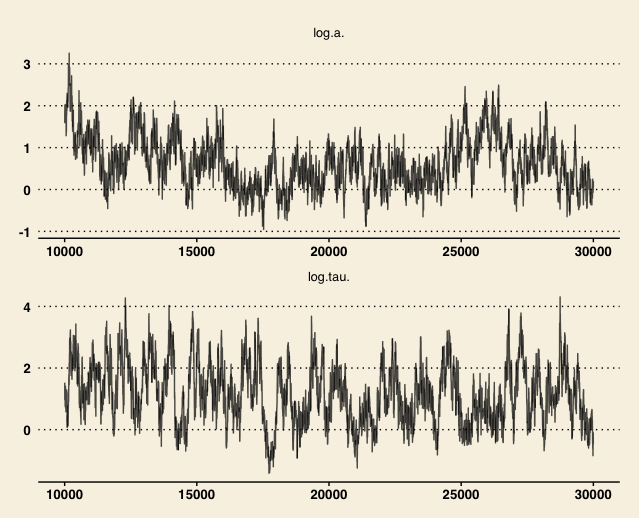}
		
		\medskip
		
		\includegraphics[width=0.85\textwidth]{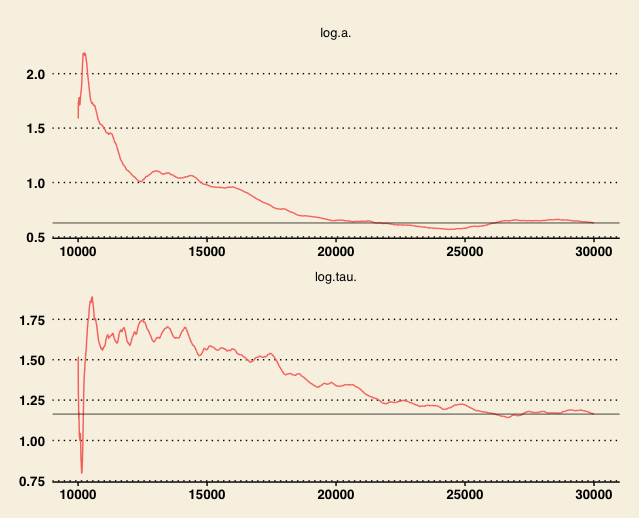}
	\end{center}
	\captionsetup{width=0.85\textwidth, font=small}
	\caption{Graphical convergence diagnostics for de-noising the Doppler function in Figure \ref{fig:doppler:noisy}. Top: Trace plots  for  logarithms of hyperparameters $a$ and $\tau_{gl}$. Bottom: running means.}
	\label{fig:hyper:trace}
\end{figure}

On the other hand, the chains for the global hyperparameters $a$ and $\tau_{gl}$ (again corresponding to level $j=3$ of DWT) are mixing somewhat slower, see  Figures \ref{fig:hyper:diag} and \ref{fig:hyper:trace}. Large amounts of data and long chain runs are required for precise learning of the hyperparameters $a$ and $\tau_{gl}$. To improve mixing of these chains, one might consider updating $a$ and $\tau_{gl}$ via the Metropolis-adjusted Langevin algorithm rather than the random walk Metropolis-Hastings algorithm.  However, it appears that even with somewhat less accurate knowledge of $a$ and $\tau_{gl}$, wavelet coefficients can still be inferred efficiently. Finally, we note that mixing of Markov chains for coarser levels of DWT is in general faster than for finer levels (plots not shown).

\section{}
\label{app:addsim}

In this appendix we present results of some additional simulations complementing those from Section~\ref{sec:simulations}. Here the sample size was $N=256.$ This is a challenging setting for the caravan method, since in general detection of structures in wavelet coefficients is easier with large sample sizes. We used two values for the signal-to-noise ratio: low $\operatorname{SNR}=3$  and high $\operatorname{SNR}=7$. We employed the $\operatorname{LA}(8)$ filter. The number of processed levels of the DWT decomposition was $J_0=4$, while that of the MODWT was  $J_0=3;$ these choices might be suboptimal, but nevertheless give an insight into relative performance of the de-noising methods. The hyperparameters of the caravan prior were set as in Sections \ref{sec:simulations} and \ref{sec:nmr}; see Appendix \ref{app:hyper} for specific values. Markov chains for all signals were run for $30\, 000$ iterations ($100\, 000$ iterations for the HeaviSine signals), with the first third of the samples discarded as a burn-in. No thinning was used.

{
	{\small
		\begin{table}
			\begin{center}
				\captionsetup{width=0.85\textwidth, font=small}
				\caption{Average square errors (over $50$ simulation runs) for various test functions and methods. The sample size is $N=256$, the $\operatorname{LA}(8)$ filter is used, and periodic boundary conditions are imposed. The number of DWT levels equals $J_0=4$.}
				\begin{tabular}{lrrrr@{\hskip 0.25in}rrrr}
					\toprule 
					& \multicolumn{4}{c}{{\bf {\hskip -0.25in} Low noise}} & \multicolumn{4}{c}{{\bf High noise}}\\
					\cmidrule(l{0.1in}r{0.35in}){2-5} \cmidrule(l{0.1in}r{0.1in}){6-9}
					{\bf Method} & {\bf bmp} & {\bf blk} & {\bf dpl} & {\bf hvs}
					& {\bf bmp} & {\bf blk} & {\bf dpl} & {\bf hvs} \\
					\midrule
					Caravan (mean)   & {\color{blue} \it 2.8} & {\color{blue} \it 2.7}     & {\color{blue} \it 1.5} & 1.1 & {\color{blue} \it 16.2} & 15.3
					& {\color{blue} \it 6.9} & {\color{blue} \it 3.1}\\
					Caravan (median) & {\color{blue} \it 2.8} & 2.8 &  {\color{blue} \it 1.5} & 1.1 & 18.0 & 16.4
					& 7.2 & {\color{blue} \it 3.1}\\
					EBayes (mean)    & 3.4 & 2.8 &  2.2 & {\color{blue} \it 1.0} & 17.4 & {\color{blue} \it 13.8}
					& 9.3 & 3.3 \\
					EBayes (median)  & 4.2 & 3.1 & 2.4 & {\color{blue} \it 1.0} & 21.5 & 15.4
					& 10.3 & {\color{blue} \it 3.1}\\
					\bottomrule
				\end{tabular}
				\label{table:dwt:3}
			\end{center}
		\end{table}
	}
}

{
	{\small 
		\begin{table}
			\begin{center}
				\captionsetup{width=0.85\textwidth, font=small}
				\caption{Average square errors (over $50$ simulation runs) for various test functions and methods. The sample size is $N=256$, the $\operatorname{LA}(8)$ filter is used, and periodic boundary conditions are imposed. The number of MODWT levels equals $J_0=3$.}
				\begin{tabular}{lrrrr@{\hskip 0.25in}rrrr}
					\toprule 
					& \multicolumn{4}{c}{{\bf {\hskip -0.25in} Low noise}} & \multicolumn{4}{c}{{\bf High noise}}\\
					\cmidrule(l{0.1in}r{0.35in}){2-5} \cmidrule(l{0.1in}r{0.1in}){6-9}
					{\bf Method} & {\bf bmp} & {\bf blk} & {\bf dpl} & {\bf hvs}
					& {\bf bmp} & {\bf blk} & {\bf dpl} & {\bf hvs} \\
					\midrule
					Caravan (mean)   & {\color{blue} \it 2.3} & {\color{blue} \it 2.3} &  {\color{blue} \it 1.3} & {\color{blue} \it 1.0}   & {\color{blue} \it 12.2} & {\color{blue} \it 13.8}
					& 6.3 & 4.2 \\
					Caravan (median) & {\color{blue} \it 2.3} & {\color{blue} \it 2.3} & {\color{blue} \it 1.3} & {\color{blue} \it 1.0} & 12.7 & 14.6
					& {\color{blue} \it 6.2} & 4.0\\
					EBayes (mean)    & 2.6 & 2.4 & 1.6 & {\color{blue} \it 1.0} &  13.0 & 14.2 & 8.8 & 3.9 \\
					EBayes (median)  & 3.0 & 2.6 & 1.6 & 1.1 & 14.5 & 16.1 & 9.3 & {\color{blue} \it 3.8} \\
					\bottomrule
				\end{tabular}
				\label{table:modwt:3}
			\end{center}
		\end{table}
	}}

We report estimation errors for the DWT de-noising (averaged over $50$ independent simulation runs) in Table~\ref{table:dwt:3}. Results for the MODWT de-noising (using the same synthetic data as in the DWT case) are given in Table~\ref{table:modwt:3}. While standard deviations are not displayed in the tables, they were circa $15-20\%$ of the estimated values. General conclusions that follow from these simulation results are similar to those already reached in Section \ref{sec:simulations}. Note that in some cases the results of the DWT de-noising are somewhat better than those for the MODWT de-noising. This may be attributable to the choice of the decomposition level $J_0$ and performance of the estimator $\hat{\sigma}_j$ of the standard deviation $\sigma_j.$ See Remark~\ref{rem:perf} in the main body of the paper. Furthermore, there is no contradiction between the fact that the estimation errors in the tables in Section \ref{sec:simulations} are larger than the corresponding ones in the tables of the present appendix: a larger sample size in Section \ref{sec:simulations} automatically implies that a larger number of parameters needs to be estimated, hence a possibility for a growth in the square error \eqref{eq:sqe}. Moreover, as in both settings the underlying `true' signals are scaled to have standard deviations equal to $1$ (see Subsection \ref{subsec:test}), the smaller sample size case cannot be viewed as a mere sub-sampling of the larger sample case.

\bibliographystyle{apa-good}
\bibliography{bibliography}

\end{document}